\documentclass[12pt]{article}
\usepackage{amsmath, amsthm, amssymb, bbm, setspace,bigints}
\usepackage{graphicx,psfrag,epsf}
\usepackage{enumerate}
\usepackage{url} 
\usepackage{caption}
\usepackage{subcaption}
\usepackage[toc,page]{appendix}
\usepackage{pdfpages}
\usepackage{epstopdf}
\usepackage{booktabs}
\usepackage{enumerate}
\usepackage{ragged2e}
\usepackage{csquotes}
\usepackage{textcomp}
\usepackage{hhline}
\usepackage{multirow}
\usepackage{xcolor,colortbl}
\usepackage{url}
\usepackage{chngcntr}
\usepackage{lipsum}
\usepackage{lmodern}
\usepackage{tcolorbox}
\usepackage{wrapfig}
\usepackage{lscape}
\usepackage{rotating}
\usepackage{algorithm}
\usepackage{algpseudocode}
\usepackage{float}

\RequirePackage[OT1]{fontenc} \RequirePackage{amsthm,amsmath}
\usepackage[colorlinks=true,linkcolor=blue,citecolor=blue,urlcolor=blue]{hyperref}
\usepackage{xr}

\usepackage{mathtools}
\usepackage[square, authoryear]{natbib}

\usepackage{etoolbox}
\usepackage{soul}
\usepackage{accents}

\def\bSig\mathbf{\Sigma}

\newcommand{\tr}{\mbox{tr}}
\newcommand{\bigCI}{\mathrel{\text{\scalebox{1.07}{$\perp\mkern-10mu\perp$}}}}

\makeatletter
\newcommand*{\sortentry}[1]{%
  \if@filesw
    \immediate\write\@auxout{\string\scNAT@aux@sortentry{#1}}%
  \fi}
\newcommand*{\scNAT@aux@sortentry}{%
  \listgadd{\scNAT@bibsortlist}}
\newcommand*{\scNAT@bibsortlist}{}

\newcommand*{\scNAT@citekeys}{}

\newcommand*{\scNAT@writetocitelistsort}[1]{%
  \ifinlist{#1}{\scNAT@citekeys}
    {\ifdefvoid{\NAT@cite@list}
       {\def\NAT@cite@list{#1}}
       {\expandafter\def\expandafter\NAT@cite@list\expandafter{\NAT@cite@list,#1}}%
     \listgadd{\scNAT@foundkeys}{#1}}
    {}}

\newcommand*{\scNAT@writetocitelistforgotten}[1]{%
  \ifinlist{#1}{\scNAT@foundkeys}
    {}
    {\ifdefvoid{\NAT@cite@list}
       {\def\NAT@cite@list{#1}}
       {\expandafter\def\expandafter\NAT@cite@list\expandafter{\NAT@cite@list,#1}}}}

\newcommand*{\scNAT@sortcites}[1]{%
  \let\NAT@cite@list\@empty
  \let\scNAT@citekeys\@empty
  \let\scNAT@foundkeys\@empty
  \forcsvlist{\listadd{\scNAT@citekeys}}{#1}%
  \forlistloop{\scNAT@writetocitelistsort}{\scNAT@bibsortlist}%
  \forlistloop{\scNAT@writetocitelistforgotten}{\scNAT@citekeys}%
}

\def\NAT@citex%
  [#1][#2]#3{%
  \NAT@reset@parser
  \NAT@sort@cites{#3}%
  \scNAT@sortcites{#3}
  \NAT@reset@citea
  \@cite{\let\NAT@nm\@empty\let\NAT@year\@empty
    \@for\@citeb:=\NAT@cite@list\do
    {\@safe@activestrue
     \edef\@citeb{\expandafter\@firstofone\@citeb\@empty}%
     \@safe@activesfalse
     \@ifundefined{b@\@citeb\@extra@b@citeb}{\@citea%
       {\reset@font\bfseries ?}\NAT@citeundefined
                 \PackageWarning{natbib}%
       {Citation `\@citeb' on page \thepage \space undefined}\def\NAT@date{}}%
     {\let\NAT@last@nm=\NAT@nm\let\NAT@last@yr=\NAT@year
      \NAT@parse{\@citeb}%
      \ifNAT@longnames\@ifundefined{bv@\@citeb\@extra@b@citeb}{%
        \let\NAT@name=\NAT@all@names
        \global\@namedef{bv@\@citeb\@extra@b@citeb}{}}{}%
      \fi
     \ifNAT@full\let\NAT@nm\NAT@all@names\else
       \let\NAT@nm\NAT@name\fi
     \ifNAT@swa\ifcase\NAT@ctype
       \if\relax\NAT@date\relax
         \@citea\NAT@hyper@{\NAT@nmfmt{\NAT@nm}\NAT@date}%
       \else
         \ifx\NAT@last@nm\NAT@nm\NAT@yrsep
            \ifx\NAT@last@yr\NAT@year
              \def\NAT@temp{{?}}%
              \ifx\NAT@temp\NAT@exlab\PackageWarningNoLine{natbib}%
               {Multiple citation on page \thepage: same authors and
               year\MessageBreak without distinguishing extra
               letter,\MessageBreak appears as question mark}\fi
              \NAT@hyper@{\NAT@exlab}%
            \else\unskip\NAT@spacechar
              \NAT@hyper@{\NAT@date}%
            \fi
         \else
           \@citea\NAT@hyper@{%
             \NAT@nmfmt{\NAT@nm}%
             \hyper@natlinkbreak{%
               \NAT@aysep\NAT@spacechar}{\@citeb\@extra@b@citeb
             }%
             \NAT@date
           }%
         \fi
       \fi
     \or\@citea\NAT@hyper@{\NAT@nmfmt{\NAT@nm}}%
     \or\@citea\NAT@hyper@{\NAT@date}%
     \or\@citea\NAT@hyper@{\NAT@alias}%
     \fi \NAT@def@citea
     \else
       \ifcase\NAT@ctype
        \if\relax\NAT@date\relax
          \@citea\NAT@hyper@{\NAT@nmfmt{\NAT@nm}}%
        \else
         \ifx\NAT@last@nm\NAT@nm\NAT@yrsep
            \ifx\NAT@last@yr\NAT@year
              \def\NAT@temp{{?}}%
              \ifx\NAT@temp\NAT@exlab\PackageWarningNoLine{natbib}%
               {Multiple citation on page \thepage: same authors and
               year\MessageBreak without distinguishing extra
               letter,\MessageBreak appears as question mark}\fi
              \NAT@hyper@{\NAT@exlab}%
            \else
              \unskip\NAT@spacechar
              \NAT@hyper@{\NAT@date}%
            \fi
         \else
           \@citea\NAT@hyper@{%
             \NAT@nmfmt{\NAT@nm}%
             \hyper@natlinkbreak{\NAT@spacechar\NAT@@open\if*#1*\else#1\NAT@spacechar\fi}%
               {\@citeb\@extra@b@citeb}%
             \NAT@date
           }%
         \fi
        \fi
       \or\@citea\NAT@hyper@{\NAT@nmfmt{\NAT@nm}}%
       \or\@citea\NAT@hyper@{\NAT@date}%
       \or\@citea\NAT@hyper@{\NAT@alias}%
       \fi
       \if\relax\NAT@date\relax
         \NAT@def@citea
       \else
         \NAT@def@citea@close
       \fi
     \fi
     }}\ifNAT@swa\else\if*#2*\else\NAT@cmt#2\fi
     \if\relax\NAT@date\relax\else\NAT@@close\fi\fi}{#1}{#2}}
\makeatother

\makeatletter
\newcommand*{\addFileDependency}[1]{
  \typeout{(#1)}
  \@addtofilelist{#1}
  \IfFileExists{#1}{}{\typeout{No file #1.}}
}
\makeatother

\newcommand*{\myexternaldocument}[1]{
    \externaldocument{#1}
    \addFileDependency{#1.tex}
    \addFileDependency{#1.aux}
}
\myexternaldocument{msgr_supp}

\newcommand{\blind}{0}

\begin{document}

\def\spacingset#1{\renewcommand{\baselinestretch}%
{#1}\small\normalsize} \spacingset{1}


\if0\blind
{
  \title{\vspace{-1cm} \bf Multi-resolution Spatial Graphical Regression Models for Hierarchical Spatial Transcriptomics Data}
  \author{
Liying Chen$^{1}$,
Satwik Acharyya$^{2}$,
Allison M.~May$^{3}$,
Aaron M.~Udager$^{4}$,\\
Evan T.~Keller$^{4,5,6,7}$, Veerabhadran Baladandayuthapani$^{1,8}$\\[0.5em]
\small $^{1}$Department of Biostatistics, University of Michigan, Ann Arbor, MI, USA\\
\small $^{2}$Department of Biostatistics, University of Alabama at Birmingham, Birmingham, AL, USA\\
\small $^{3}$Department of Urology, University of Virginia, Charlottesville, VA, USA\\
\small $^{4}$Department of Pathology, University of Michigan, Ann Arbor, MI, USA \\
\small $^{5}$Department of Urology, University of Michigan, Ann Arbor, MI, USA\\
\small $^{6}$Single Cell Spatial Analysis Program, University of Michigan, Ann Arbor, MI, USA\\
\small $^{7}$Biointerfaces Institute, University of Michigan, Ann Arbor, MI, USA \\
\small $^{8}$ Corresponding author: veerab@umich.edu
}
  \maketitle
} \fi

\if1\blind
{
  \bigskip
  \bigskip
  \bigskip
  \begin{center}
    {\Large\bf
    Multi-resolution Spatial Graphical Regression Models for Hierarchical Spatial Transcriptomics Data}
\end{center}
  \medskip
} \fi

\bigskip

\vspace{-1cm}
\begin{abstract}
Advances in spatial transcriptomics (ST) technologies enable systematic molecular characterization of tumor microenvironment, tumor gradients and gene regulatory networks. Cancer progression is known to vary along pathological gradients, yet existing network approaches for gene network inference typically ignore hierarchical spatial organization across the tumor. We develop a Bayesian multi-resolution spatial graphical regression (\texttt{mSGR}) framework to infer spatially varying gene networks from multi-resolution ST data. The proposed model allows precision matrices to vary across hierarchically structured spatial domains, capturing both local and global organization within the tumor. To identify spatially varying regulatory relationships, we introduce a spatially structured edge selection strategy that borrows strength across regions according to spatial proximity and pathological gradients, while Gaussian-process priors flexibly model spatial variation in edge strengths. Scalable inference is achieved through an augmented mean-field variational Bayes algorithm with node-wise parallel regressions, enabling efficient estimation in high-dimensional settings. Simulation studies demonstrate improved recovery of network structures compared with competing approaches. Applying \texttt{mSGR} to multi-resolution ST data from kidney cancer reveals stronger regulatory connectivity in transitional regions of epithelial-mesenchymal transition pathway and identifies hub genes along the tumor gradient, illustrating how spatially resolved network analysis can provide key insights into tumor microenvironment organization.
\end{abstract}

\noindent%
\vfill



\newpage
\spacingset{1.8} 
\section{Introduction}\label{sec:intro}\vspace{-1.1em}
\noindent \textbf{Scientific background and rationale}:
The tumor microenvironment (TME), an ecosystem consisting of cancer, immune, and stromal cells, is considered a critical factor influencing tumorigenesis, tumor spread, metastasis, and treatment response \citep{larson2025tumor, bilotta2022managing}.  Systematic investigation of the TME is crucial for understanding the complex interplay between cancer cells and immune cells in their surrounding spatial regions, such as stromal, immune, and vascular components. The spatial organization of cells within the TME is fundamental to cancer biology, which drive key oncogenic processes such as tumor progression, immune evasion, and therapeutic resistance \citep{marusyk2020intratumor, seferbekova2023spatial}. Traditional bulk and single-cell sequencing techniques lack spatial information, limiting their ability to capture the complex architecture of the TME. In contrast, spatial transcriptomics (ST) has emerged as a transformative technology that bridges this gap by enabling high-resolution mapping of gene expression directly within spatial tissue architecture. These advances provide a powerful framework for characterizing tissue heterogeneity and for the study of key signaling pathways leading to novel spatial biomarker identifications \citep{dong2022deciphering, sun2020statistical}. Modern ST technologies such as CosMx\texttrademark{} offer cellular resolution, and researchers often adopt a field-of-view (FOV)–based design, selecting representative spatial regions across distinct tumor areas. This approach enables high-resolution characterization of intra-tumor heterogeneity and spatial gradients within the TME.

\begin{figure}[ht!]
  \centering
  \includegraphics[scale=0.7]{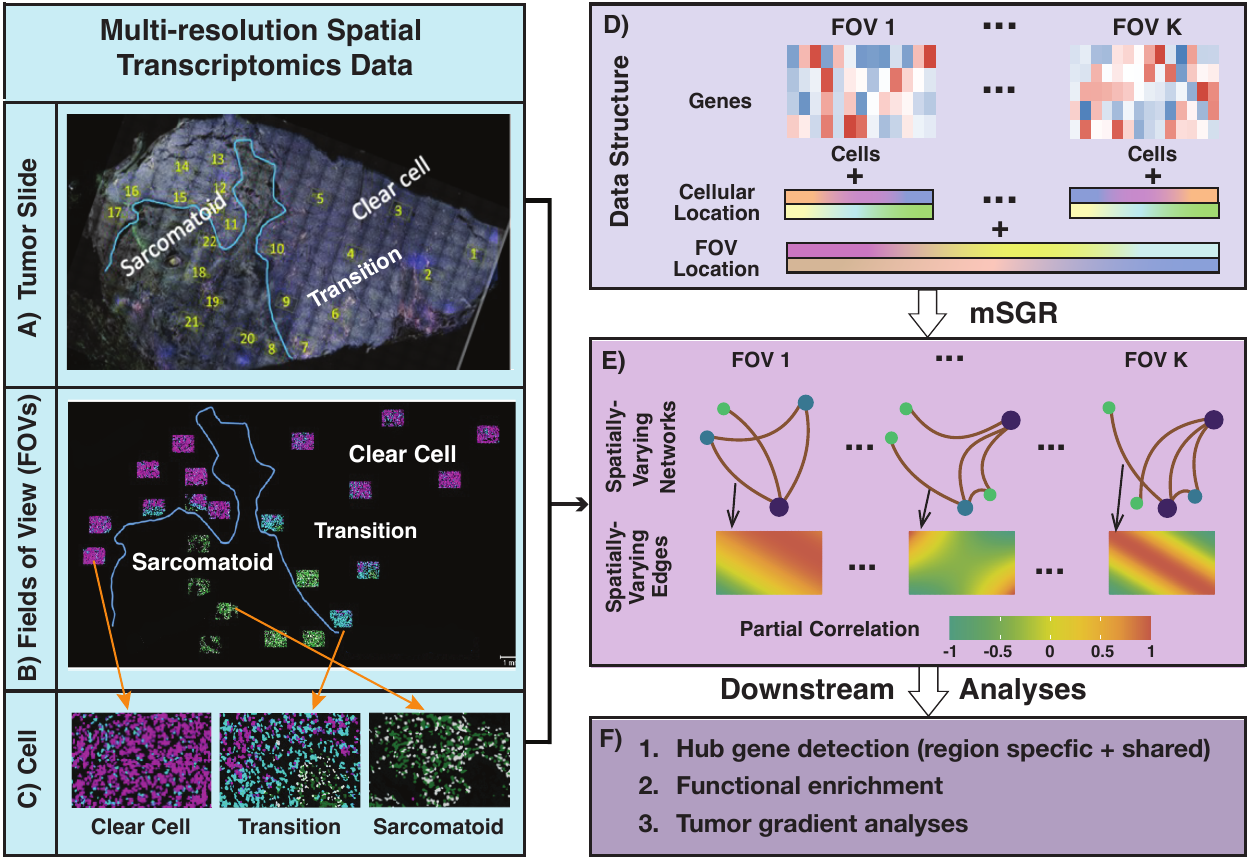}
  \vspace{-0.5em}
  \caption{\small \textbf{Overview of the Multi-resolution Spatial Graphical Regression (\texttt{mSGR}) framework:} Tumor slide in Figure \ref{fig:mSGR_overview}A is partitioned into multiple FOVs, each containing spatially resolved information on (i) FOV location within the tissue slide at a macro level (Figure \ref{fig:mSGR_overview}B), and (ii) spatial coordinates of single cells at a micro level (Figure \ref{fig:mSGR_overview}C). We implement the \texttt{mSGR} framework on hierarchical spatial gene expression data structure (Figure \ref{fig:mSGR_overview}D) to infer multi-resolution spatially varying graphs to capture regional heterogeneity across the tissue (Figure \ref{fig:mSGR_overview}E). In the downstream analyses on Figure \ref{fig:mSGR_overview}F, we focus on hub gene identification (both region-specific and shared), functional enrichment of key pathways, and tumor gradient analyses across clear cell, transition, and sarcomatoid regions.}
  \label{fig:mSGR_overview}
  \vspace{-1em}
\end{figure}

\noindent \textbf{Multi-resolution ST in renal cancer}: The motivating data for this work arise from a novel ST-based study in renal (kidney) cancer (see Figure \ref{fig:mSGR_overview} left panels). Specifically, a sarcomatoid renal cell carcinoma (sRCC) tissue obtained from a patient was profiled at multiple FOVs across the tumor using an imaging-based ST platform, CosMx\texttrademark{} Spatial Molecular Imager, to generate high throughput multi‑resolution spatial gene expression ($n=960$ genes) at the single-cell level (Figure \ref{fig:mSGR_overview}A). At the ``macro" level in Figure \ref{fig:mSGR_overview}B, the FOVs capture a {\it tumor gradient} that was annotated by two genitourinary pathologists into three histopathological subtypes: sarcomatoid, clear cell, and transition. Sarcomatoid regions display aggressive, dedifferentiated phenotypes with loss of epithelial features and mesenchymal morphology \citep{delahunt2013international}, driven by epithelial–mesenchymal transition (EMT) programs \citep{bostrom2012sarcomatoid, vcugura2024epithelial}. In contrast, clear cell regions retain epithelial characteristics and are generally less aggressive \citep{may2024association} with hypothesized transition regions being somehwere in between. At the ``micro" cellular level within each of the ($n=23$) FOVs (\ref{fig:mSGR_overview}C), single cells exhibit local niche specific gene expression dynamics (range: 399 to 1705; total ~26K cells). This nested micro within macro architecture of tumors engenders the discovery of spatially varying regulatory biological mechanisms within local niches and global tumor gradients \citep{marusyk2020intratumor}. 

\noindent \textbf{Scientific problem}: Several computational methods have been developed for ST data, primarily focusing on the identification of spatially variable genes \citep{sun2020statistical, yan2025categorization} and spatial regions \citep{zhao2021spatial,dong2022deciphering}. However, the inference of spatially varying genomic networks remains understudied. Genomic networks provide a system-level representation of molecular regulation, where nodes correspond to genes and edges encode some form of association or dependence. Modeling spatially varying gene networks enables the study of gene association patterns across different spatial regions in the TME. Network or graph-based perspectives are essential for studying regulatory mechanisms that are not captured by single-gene analyzes, thereby offering deeper insights into the organizational principles of spatial gene expression dynamics \citep{wang2025spclue}.

In this paper, we aim to incorporate the spatial hierarchical data structure at both micro (within-FOV) and macro (between-FOV) scales (Figure \ref{fig:mSGR_overview}D) and infer multi-resolution spatially varying graphs (Figure \ref{fig:mSGR_overview}E). By leveraging spatial information from FOV locations along the tumor gradient, our framework provides a parsimonious representation of spatial network structures. These network summary measures can be used for downstream analyzes such as hub-gene identification and functional pathway enrichment (Figure \ref{fig:mSGR_overview}F). 
\noindent  \textbf{Existing methods and limitations}: Recent efforts have begun addressing spatial heterogeneity in network estimation for ST data. Neural-network–based algorithmic approaches integrate spatial adjacency or embedding information to infer gene–gene associations \citep{yuan2020gcng,xu2021costa}. In contrast, modeling-based approaches represent spatial variation through probabilistic formulations of the underlying dependence structures. Broadly, these modeling-based methods can be classified into two categories in terms of modeling spatially varying marginal and conditional dependence through covariance and precision matrices, respectively. The marginal correlation modeling approaches estimate the joint structure through parametric or nonparametric covariance functions, often leveraging spatial kernels or decomposition techniques to account for spatial heterogeneity \citep{acharyya2022spacex, bernstein2022spatialcorr,chakrabarti2024joint,Vasconcelos2025.07.23.666450}. The conditional dependence–based approaches extend Gaussian graphical models (GGMs) to a spatial setting using spatially varying coefficient frameworks \citep{chen2025probabilistic,acharyya2025spatially}. The conditional approaches yield sparser and biologically interpretable networks while mitigating the confounding effects from indirect associations. However, all of these methods are \ul{limited} to analyzing a single spatial tissue domain or assume independence across samples in the case of multi-sample analyses. 

In non-spatial settings, several approaches have been proposed for graphical model estimation for dependent samples, termed multiple GGMs (mGGMs), that jointly estimate a collection of precision matrices, enabling the characterization of shared and view specific conditional structures. Joint graphical lasso-based methods leverage structural similarity while accounting for group specific variation \citep{guo2011joint,danaher2014joint}. Bayesian approaches for mGGMs link multiple graph structures through a Markov random field (MRF) prior \citep{peterson2015bayesian}, spike and slab formulations \citep{li2019bayesian,yang2021gembag}, or graphical horseshoe priors \citep{lingjaerde2024scalable,Busatto21012025} to accord network sparsity and heterogeneity.
However, none of these methods incorporate spatial dependency or allow for the inference of spatially varying graphical structures. In summary, all of these aforementioned classes of methods are unable to infer spatially varying genomic graphs with a hierarchical multi-resolution spatial structure, as shown in Figure \ref{fig:mSGR_overview}, incorporating both macro- and micro-level tumor information.  
\noindent \textbf{Novel contributions}: To address these challenges, we propose a multi-resolution spatial graphical regression (\texttt{mSGR}) -- a Bayesian framework to infer spatially varying graphs across hierarchical spatial domains. Our model generalizes the mGGMs, conferring several key advantages. First, it flexibly models the precision matrix through Gaussian process priors, allowing the underlying graphical structure to vary across hierarchical spatial structures. Second, leveraging multi-resolution ST, the model borrows information across FOVs employing a Bayesian structured variable selection technique, where the edge-inclusion indicators incorporate the underlying topological spatial domain. This serves as a scalable alternative to MRF and structured spike-and-slab priors and allows for symmetry and positive definiteness constraints. Third, we impose a sparsity constraint to mimic the sparse nature of ST data, which enhances statistical interpretability and computational tractability. Finally, we propose a novel augmented variational inference algorithm to fit the \texttt{mSGR} model, which allows for parallelization and provides a scalable and efficient alternative to computationally intensive Markov Chain Monte Carlo (MCMC) methods for fitting such large-scale spatial datasets.  


Comprehensive simulation studies demonstrate that the proposed \texttt{mSGR} framework accurately recovers spatially varying network structures across diverse settings. By jointly modeling multi-resolution spatial dependencies and spatially adaptive information sharing across FOVs, \texttt{mSGR} attains higher accuracy than many existing methods, underscoring its robustness and effectiveness in characterizing spatial heterogeneity. Analyses of sRCC ST data using \texttt{mSGR} reveal FOV-specific network structures of EMT genes, capturing spatially varying conditional dependencies between biologically linked gene pairs, such as collagen-encoding genes, and the dynamic hub behavior of key regulators such as MALAT1 and VIM along the tumor progression gradient. Furthermore, \texttt{mSGR} identifies markedly higher network connectivity in sarcomatoid regions, particularly within pathways associated with invasion, metastasis, and resistance to cell death, delineating the mechanistic rewiring accompanying aggressive tumor phenotypes.

\noindent \textbf{Sectional contents}: The remainder of this paper is organized as follows. Section \ref{sec:mSGM} introduces the proposed \texttt{mSGR} framework along with the model formulation, characterization, and prior specifications. The details of variational inference of \texttt{mSGR} are provided in Section \ref{sec:mfvb}. Section \ref{sec:simStudy} presents comprehensive simulation studies to assess the performance of \texttt{mSGR} under different levels of spatial heterogeneity, data dimensions, and comparisons with competing methods. Section \ref{sec:Application} applies \texttt{mSGR} to the multi-resolution ST from RCC tissue, where we study multi-resolution spatially varying graphs to infer spatial regulatory genes. We conclude with a discussion of methodological extensions and potential biomedical implications in Section \ref{sec:Discussion}. The \texttt{mSGR} R package is open-source and available on GitHub at https://github.com/***.



\vspace{-2em}

\section{Multi-resolution spatial graphical models}\label{sec:mSGM} \vspace{-1em}
We start by establishing the notational details for the multi-resolution ST data structure. Suppose a tissue slide is composed of $K$ FOVs which are regions of interest where the $k$-th $\{ k \in (1, \dots, K) \}$ FOV consists of $N_k$ spatially indexed cells. The FOV specific spatial surface is denoted as $\mathbb{S}^k = \{s_{n_k}^k\}_{n_k = 1}^{N_k} \subseteq \mathbb{R}^2$ (micro-resolution) and the whole spatial domain of the tissue slide can be represented as $\mathbb{S} = \bigcup_{k=1}^K \mathbb{S}^k$ (macro-resolution). The observed gene expression matrix from $p$ genes in the $k$th FOV is denoted as $\mathbf{Y}(\mathbb{S}^k) = [\mathbf{Y}_{1}^\top(\mathbb{S}^k), \dots, \mathbf{Y}_{p}^\top(\mathbb{S}^k)] \in \mathbb{R}^{N_k \times p}$. 

Our inferential object is multi-resolution spatially varying graphs, which are characterized by $G^{k}[V,E^{k}(\mathbb{S}^k)]$; where 
$V$ denotes the vertex set of $p$ genes and 
$E^{k}(\mathbb{S}^k)$ denotes spatially varying edge set across $K$ FOVs encoding spatially varying conditional dependencies. Our primary inferential goal is the spatially varying graph topology while borrowing information on $E^{k}(\mathbb{S}^k)$ across FOVs. To elucidate the core construction, we first start with mGGM in Section \ref{subsec:mGGM} and generalize to spatial settings to introduce spatial mGGMs and \texttt{mSGR} in Section \ref{subsec:Spatial_mGGM} and Section \ref{subsec:mSGR} respectively.

\noindent \textbf{Multiple Gaussian graphical models}: \label{subsec:mGGM} 
Traditional GGMs encode conditional dependencies among variables through non-zero entries of the precision matrix \citep{dempster1972covariance}. A standard (nonspatial) GGM is represented by an undirected graph $G = (V, E)$, where edges represent non-zero partial correlations under a multivariate Gaussian model $ \textbf{Y} \sim \mathcal{N}(0, \boldsymbol{\Omega}^{-1}) $, with precision matrix $\boldsymbol{\Omega}$. Two nodes $i$ and $j$ are connected if their partial correlation $\rho_{ij} = \text{Corr}(\textbf{Y}_i, \textbf{Y}_j \mid \textbf{Y}_l, l \ne i,j)$ is non-zero.

In mGMM settings, the observed data is sourced from $K$ related views with $\textbf{Y}^{k} \sim N(0,\boldsymbol{\Omega}^{-1}_{k})$ \citep{lingjaerde2024scalable}. For each $k$, the inferential object is a view-specific precision matrix $\boldsymbol{\Omega}_{k} = ( \omega_{ij}^{k} )_{p \times p}$, corresponding to a graph $G^{k} = (V, E^{k})$. The goal is to jointly infer $\{ \boldsymbol{\Omega}_{k} \}_{k=1}^K$, capturing both shared and view-specific graph structures. The $k$-th view specific partial correlation between node $i$ and $j$ is quantified as $\rho_{ij}^{k} = \text{Corr}(\textbf{Y}_{i}^{k}, \textbf{Y}_{j}^{k} \mid \textbf{Y}_{l}^{k}, l \ne i,j) = -\omega_{ij}^{k} / \sqrt{\omega_{ii}^{k} \omega_{jj}^{k}}$. This implies that $\rho_{ij}^{k} \ne 0$ (or equivalently $\omega_{ij}^{k} \ne 0$) indicates conditional dependence between nodes $i$ and $j$ in the $k$-th graph. The graph structure can be represented with a collection of undirected graphs $G^{k} = (V, E^{k})$, where $V = \{1, \dots, p\}$, $E^{k} = \{(i,j): \rho_{ij}^{k} \ne 0, \, i \ne j\}$. The joint graph estimation in mGGM setting reduces to identifying the non-zero elements of the precision matrices while borrowing strengths across the related views. However, these models impose a single shared graph for all cells within a given view i.e. assume independence between cells. Hence, we generalize mGGMs to accommodate spatial dependencies among cells within a given view (FOV in our case). 

\vspace{-1.5em}
\subsection{Multi-resolution spatially varying GGM}\label{subsec:Spatial_mGGM}\vspace{-0.5em}
We introduce multiresolution spatial GGM which infers spatially varying genomic graphs while leveraging structured information across FOVs. 
We consider the joint distribution of $\textbf{Y}(s^k) = [y_{1}(s^{k}),\dots,y_{p}(s^{k}) ]$ as \vspace{-1.5em}
\begin{equation*}\label{eq:Spatial_mGGM}
\textbf{Y}(s^k) \sim \mathcal{N}\left(0, \boldsymbol{\Omega}^{-1}(s^k)\right), \hspace{0.1cm} \forall \hspace{0.1cm} s^k \in \mathbb{S}^k. 
\vspace{-1.5em}
\end{equation*}
where $\boldsymbol{\Omega}(s^k) = \{ \omega_{ij}(s^{k}) \}_{p \times p}$ is the spatially varying precision matrices across FOVs. Analogously to mGGM settings, the spatially varying partial correlation across FOVs is defined  
\vspace{-1.5em}
\begin{equation*}
   \rho_{ij}(s^{k}) = \text{Corr}(\textbf{Y}_{i}(s^{k}), \textbf{Y}_{j}(s^{k}) \mid \textbf{Y}_{l}(s^{k}), l \ne i,j) = -{\omega_{ij}(s^{k})}/{\sqrt{\omega_{ii}(s^{k})\omega_{jj}(s^{k})}}. 
\vspace{-0.75em}
\end{equation*}


\noindent The corresponding multi-resolution spatial graph at location $s^{k}$ is then denoted as, 
\vspace{-1.5em}
\begin{equation*}
G^{k}(s^{k}) = [V,E^{k}(s^{k}) \hspace{0.25cm} \mbox{s.t.} \hspace{0.25cm} E^{k}(s^{k}) = \{ (i,j): i \ne j, \hspace{0.25cm} \mbox{s.t.} \hspace{0.25cm} \omega_{ij}(s^{k}) \ne 0 \}],\quad   \mbox{for any}\quad s^{k} \in \mathbb{S}^k. 
\vspace{-1.5em}
\end{equation*}
Following the notions of mGGM, the conditional independence between two nodes given all other nodes at location $s^{k}$ i.e. $\textbf{Y}_{i}(s^{k}) \bigCI \textbf{Y}_{j}(s^{k}) \mid \textbf{Y}_{l}(s^{k})$ s.t. $l \ne \{i,j\}$ is noted by $\omega_{ij}(s^{k}) = 0$. The conditional dependence of spatial graph $G^{k}(s^{k})$ structure can be characterized through non-zero elements of spatially varying precision matrices $\boldsymbol{\Omega}(s^k)$. The goal of this paper is to infer the non-zero elements of $\boldsymbol{\Omega}(s^k)$ while borrowing information across the $K$ FOVs. Next, we introduce the \texttt{mSGR} framework to achieve our inferential goal in a parsimonious manner.



\vspace{-1.5em}
\subsection{Multiresolution spatial graphical regression}\label{subsec:mSGR} \vspace{-0.75em}

The spatially varying precision matrices $\{\boldsymbol{\Omega}(\mathbb{S}^k) \}_{k=1}^{K}$ is a high dimensional object where the number of estimands are on the order of $O(\sum_{k=1}^{K}N_{k}p^{2})$. For example, for our case study where the number of cells are on the order of  $10^{4}$ and even with moderate number of genes ($p=27 \sim 115$) 
this can exceed $10^{6}$ parameters -- necessitating a computationally tractable estimation strategy. 
For example, a joint estimation strategy based on  MRF \citep{peterson2015bayesian} or Wishart priors \citep{Lenkoski01012011} becomes infeasible due computational complexity. To this end, we employ a neighborhood selection and regression-based approach \citep{meinshausen2006high,ni2019bayesian,ha2021bayesian, zhang2022high} to estimate the non-zero elements of $\{\boldsymbol{\Omega}(\mathbb{S}^k) \}_{k=1}^{K}$. 

Specifically, we introduce the \texttt{mSGR} model as 
\vspace{-1em} 
\begin{equation*}\label{eq:mSGR}
    \textbf{Y}_{i}(\mathbb{S}^k) = \sum_{j\neq i}^p \textbf{Y}_{j}(\mathbb{S}^k) \gamma_{ij}(\mathbb{S}^k) + \epsilon_i(s^k), \hspace{1em} \epsilon_i(s^k) \sim N\left(0,1/\omega_{ii}(s^k)\right) 
\vspace{-1em}
\end{equation*}
where $\gamma_{ij}(\mathbb{S}^k)$ is interpreted as the spatial graphical regression coefficients between genes $i$ and $j$ within $k$th FOV. Since our primary focus is on estimating the off diagonal elements, we assume that diagonal variances $\omega_{ii}(s^k)$ does not vary with spatial location and can be represented by a constant value $\omega_{ii}^k$.
For any location $s^k \in \mathbb{S}^k$, $\gamma_{ij}(s^k)$ is characterized as $\gamma_{ij}(s^k) = -\omega_{ij}(s^k)/\omega_{ii}^k = \rho_{ij}(s^{k})\sqrt{\omega_{jj}^k/\omega_{ii}^k}$ iff $\epsilon_{i} \bigCI Y_{-i}(s^{k})$.  This establishes one-to-one correspondence between spatial graphical regression coefficients and conditional dependencies encoded in the spatially varying precision matrix $\Omega(s^k)$. 

To enable flexibility and sparsity in the estimation of the spatial genomics networks, we decouple the spatial graphical regression coefficients $\gamma_{ij}(\mathbb{S}^k)$ as follows:
\vspace{-1em} 
\begin{equation}\label{eq:gamma_decompose}
    \gamma_{ij}(\mathbb{S}^k) = \beta_{ij}(\mathbb{S}^k) \delta_{ij}^k.  
\vspace{-1em}
\end{equation}
where $\beta_{ij}(\mathbb{S}^k)$ is the magnitude of the spatial surface along with a selection indicator $\delta^{k}_{ij}$.  
\noindent \textbf{Spatial surface modeling via Gaussian process}:
To admit a flexible spatial structure, we model $\beta_{ij}(\mathbb{S}^k)$ with nonparametric Gaussian process prior with modified squared exponential kernel \citep{shi2015thresholded,acharyya2025spatially}, enabling it to capture spatial variation within the FOVs. Using the Karhunen–Loève (KL) expansion, $\beta_{ij}(\mathbb{S}^k)$ can be approximated by a set of basis functions $B_l(\cdot)$ and corresponding coefficients $u_{ij}^k$, shown as:
\vspace{-1.25em} 
\begin{equation}
    \beta_{ij}(s^k) = \sum_{l=1}^ L u^k_{ijl} B_l(s^k) \label{eq:gp_main}.
    \vspace{-0.5em} 
\end{equation}
 Further details regarding Gaussian process-based modeling of $\beta_{ij}(\mathbb{S}^k)$ is deferred to Section \ref{suppsec:method_gp} of the Supplementary Materials.


\vspace{-1.5em}
\subsection{Bayesian structured edge selection across spatial surfaces} \label{subsec:method_selectionprior}
\vspace{-0.75em}

Our primary interest lies in modeling of selection indicator $\delta_{ij}^k$, which represents the existence of a (spatial) edge between gene $i$ and gene $j$ in FOV $k$. Under the standard formulation, the indicators $\delta$ are assumed to be independent across spatial surfaces \citep{mitchell1988bayesian}. 
However, in many applications, particularly those involving spatial, temporal, or network-structured data—selection, indicators often exhibit an underlying topographic or spatial structure, and the independence assumption is no longer hold \citep{andersen2017bayesian}. It is reasonable to expect that the joint inclusion of two selection indicators, $\delta_i$ and $\delta_j$, with spatial locations $s_i$ and $s_j$, depends on their proximity, such that $\mathbb{P}(\delta_i = 1, \delta_j = 1)$ increases as $|s_i - s_j|$ decreases. This indicates variables with spatial proximity of each other are more likely to be jointly selected or excluded. This is particularly relevant in spatial omics, where genes co-expressed within the same tumor microenvironment often exhibit correlated regulation \citep{du2024spatial, wang2025stmodule}.

To accommodate such structured dependencies, several extensions of the spike-and-slab framework have been proposed. These include the use of logistic regression–based product priors to capture group-specific effects \citep{stingo2010bayesian, menacher2024bayesian}, Ising priors to encode spatial or graph-based similarities \citep{li2010bayesian, peterson2015bayesian}, and structured spike-and-slab priors that introduce smooth spatial correlations among inclusion probabilities \citep{andersen2014bayesian, andersen2017bayesian}.


Building on these ideas, we propose a novel Bayesian structured edge selection prior on the binary inclusion variables, following the framework of \citep{andersen2014bayesian, mohammed2021tumor}.
Specifically, each indicator $\delta_{\bullet}^k$ controls the inclusion of edge between nodes over the spatial domain $\mathbb{S}^k$. These indicators are modeled as Bernoulli random variables with selection probabilities determined through a probit link, 
\vspace{-1.5em} 
\begin{equation*} 
    \delta_{ij}^k \sim Ber(\Phi(\lambda_{ij}^k)), 
\vspace{-2em}
\end{equation*}
where $\Phi(\cdot)$ denotes the standard normal cumulative distribution function. The latent variables ($\lambda_i$'s) for each node-specific regression are collected into a matrix, and are assigned a matrix-variate Gaussian prior: 
\vspace{-1em}
\begin{equation} \label{eq:matrix_normal}
\Lambda_i \in \mathbb{R}^{K \times (p-1) }  \sim \mathcal{MN}(M_i, U_i, V_i) 
\vspace{-1em}
\end{equation}



\paragraph{Characterization of $V$:} \hspace{-1em} The column covariance matrix $V_i \in \mathbb{R}^{K \times K}$ captures the macro-level spatial correlations across FOVs. Each entry, denoted as $V_i[k,k']$ reflects the prior correlation between FOVs $k$ and $k'$, enabling the model to borrow strength across FOV level resolution. This structure aligns with multi-resolution ST, since neighboring FOVs are expected to exhibit similar biological processes. In our application, we parameterize $V_i$ using a block-wise correlation structure. Specifically, FOVs belonging to different pathological regions (e.g., clear-cell, transitional, or sarcomatoid) are assumed to be independent, such that their cross-region correlations are set to zero. Within each region, the correlation between any two FOVs $k$ and $k'$ is defined as
\vspace{-1.5em}
\begin{equation*}
  V_i[k,k'] = \rho^{d_{kk'}},
  \label{eq:Vi_spatial}
  \vspace{-1.5em}
\end{equation*}
where $d_{kk'}$ denotes the spatial distance between the centroids of FOVs $k$ and $k'$, and $\rho \in (0,1)$ controls the rate of spatial decay. 
\vspace{-1.25em}
\paragraph{Characterization of $U$:} \hspace{-1em} The row covariance matrix $U_i \in \mathbb{R}^{(p-1) \times (p-1)}$ encodes the dependencies among predictor genes while regressing on the $i$-th node. This covariance structure enables incorporation of prior knowledge of dependence between genes and coherent edge selection among biologically related genes in a pathway across spatial locations. 


\vspace{-1.25em}
\paragraph{Characterization of $M$:} \hspace{-1em} The mean matrix is denoted by $M_i \in \mathbb{R}^{K \times (p-1)}$, where each row corresponds to a specific field of view (FOV) and each column corresponds to one of the $p-1$ potential predictor genes. This structure allows the model to incorporate prior knowledge about edge inclusion across spatial regions, and to capture both global and local regulatory signals.

\begin{figure}[ht!]
  \centering
  \includegraphics[width=\textwidth]{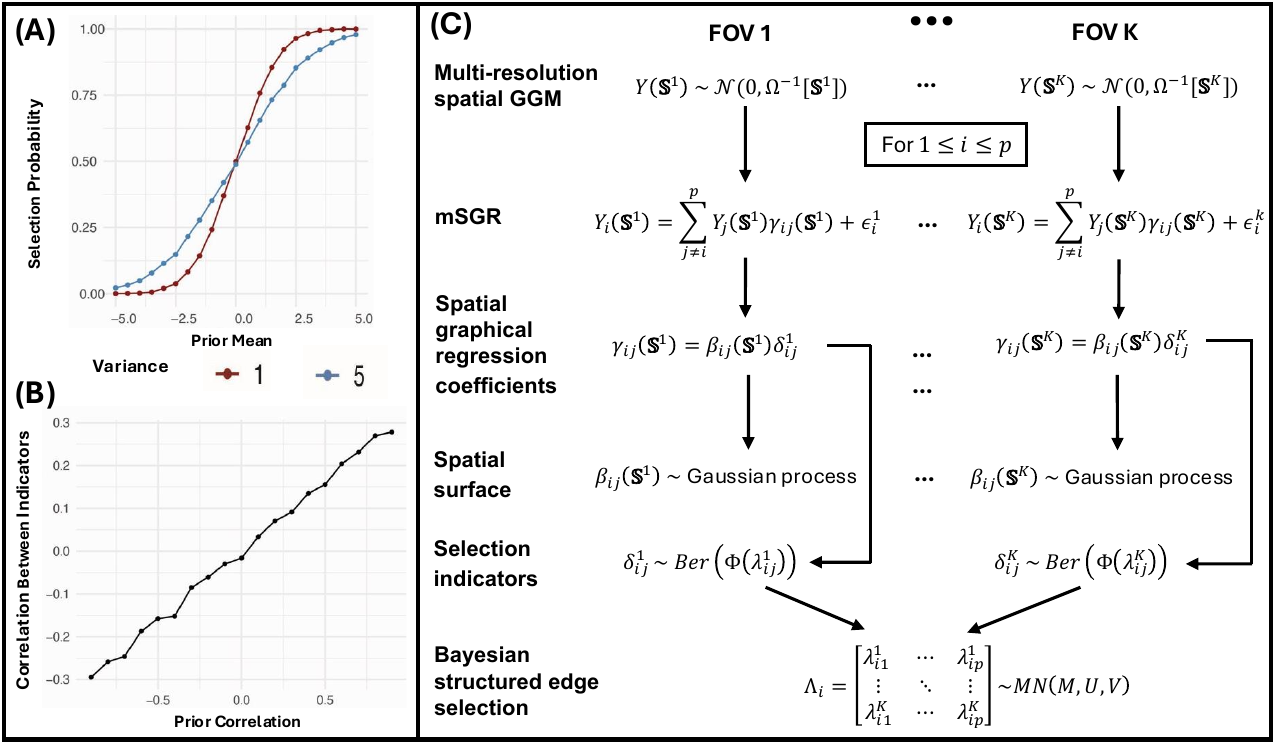}
  \caption{\small \textbf{Overview of the Bayesian multiresolution spatial graphical regression (\texttt{mSGR}) model.} \textbf{(A)} The impact of the prior mean on the marginal probability of selection for two variables with differing prior variances (1 in dark red, 5 in steel blue). As the prior mean increases, the probability of selection increases monotonically for both variables, with the higher-variance variable showing a more gradual change.
\textbf{(B)} The influence of prior correlation on the empirical correlation between binary selection indicators. As the prior correlation increases, the resulting correlation between indicators also increases. \textbf{(C)} Flow diagram of the \texttt{mSGR} model.}
  \label{fig:prior_selection}
  \vspace{-1em}
\end{figure}

\noindent In summary, the matrix normal prior $\Lambda_i \sim \mathcal{MN}(M_i, U_i, V_i)$ provides a structured and interpretable way to borrow strength and model prior beliefs about both spatial and gene-wise dependencies in the inclusion mechanism for edges in the graphical model. 
More specifically, let $\tilde{\lambda}_r$ denote the $r$-th element of $\operatorname{vec}(\Lambda_i)$, with the corresponding selection indicator being $\tilde{\delta}_r$. Under the matrix-variate Gaussian prior, the marginal distribution of $\tilde{\lambda}_r$ is Gaussian with mean $\tilde{m}_r$, the $r$-th element of $\operatorname{vec}(M_i)$, and variance $\tilde{\sigma}^2_r$, the $r$-th diagonal entry of $V_i \otimes U_i$. From equations \ref{eq:matrix_normal}, the marginal prior distribution for $\tilde{\delta}_r$ being selected is: 
\vspace{-1.5em}
\begin{equation} \label{eq:marginal_prior_delta}
    Pr(\tilde{\delta}_r = 1) = \int p(\tilde{\delta}_r = 1|\tilde{\lambda}_r)p(\tilde{\lambda}_r)d\tilde{\lambda}_r =\Phi\left(\frac{\tilde{m}_r}{\sqrt{\tilde{\sigma}^2_r+1}}\right).
\vspace{-1em}
\end{equation}
Equation \ref{eq:marginal_prior_delta} implies that the prior distribution for $\delta_{ij}^k$ is non-informative when $m_{ij}^k = 0$ with the corresponding $Pr(\delta_{ij}^k) = 0.5$, while $m_{ij}^k < 0$ or $m_{ij}^k > 0$ favor $\delta_{ij}^k =1$ or $\delta_{ij}^k =0$, respectively. Additionally, the marginal prior for $\tilde{\delta}_r$ and $\tilde{\delta}_{r'}$ is simultaneously selected can be expressed as, 
\vspace{-2em}
\begin{equation*}
    Pr(\tilde{\delta}_r = 1, \tilde{\delta}_{r'} = 1) = \int \Phi(\tilde{\lambda}_r) \ \Phi(\tilde{\lambda}_{r'}) \ \mathcal{N}\left(\operatorname{vec}(\Lambda_i)\right)d\Lambda_i,
\vspace{-1em}
\end{equation*}
which captures the dependency in prior selection probabilities induced by the joint distribution of the latent variables and the ability of the model to account for correlation or shared structure in the inclusion. This is further exemplified in Figure \ref{fig:prior_selection}A, which shows that higher prior means increase selection probabilities, while larger variances make this effect more gradual by reducing the influence of the mean. Similarly, a higher prior correlation among the latent Gaussian variables induces stronger concordance among the corresponding selection indicators (Figure \ref{fig:prior_selection}B). In practice, this induced concordance is typically smaller than the specified prior correlation due to the nonlinear thresholding transformation from latent variables to binary indicators. Nevertheless, this mechanism effectively captures the spatial dependence among FOVs, as nearby FOVs with higher prior correlation are more likely to exhibit similar inclusion patterns. Consequently, the model encourages spatially coherent variable selection while allowing localized deviations supported by the data.




\noindent \textbf{\texttt{mSGR} model overview}: Figure \ref{fig:prior_selection}C summarizes the various components of the proposed \texttt{mSGR} model. In summary, for each FOV $k$, the conditional dependence among genes is modeled through spatially graphical regression, with coefficients $\gamma_{ij}(\mathbb{S}^k) = \beta_{ij}(\mathbb{S}^k)\delta_{ij}^k$. The continuous effects $\beta_{ij}(\mathbb{S}^k)$ follow Gaussian process priors to capture smooth spatial variation. The binary indicators $\delta_{ij}^k$ indicates the presence of an edge between genes $i$ and $j$ in FOV $k$.  $\delta_{ij}^k$ are assumed to follow $\mathrm{Bernoulli}\{\Phi(\lambda_{ij}^k)\}$, which are linked through a matrix-variate normal prior on latent variables $\Lambda_i \sim \mathrm{MN}(M, U, V)$ to induce structured dependence across genes and FOVs. To complete prior specification, we assume a conjugate Gamma prior for the precision parameter, $\omega_{ii}^k \sim \text{Gamma}(a_w, b_w)$.

\vspace{-2em}
\section{Augmented variational inference for \texttt{mSGR}} \label{sec:mfvb}
\vspace{-1em}


Variational inference (VI) \citep{jordan1999introduction} offers a scalable alternative to MCMC by reframing Bayesian inference as an optimization problem \citep{blei2017variational}. Instead of drawing samples from the posterior, VI approximates the posterior distribution through a target distribution by minimizing in-between Kullback–Leibler (KL) divergence. A standard choice of target distribution is the mean-field variational family \citep{wainwright2008graphical}, which assumes a fully factorized form over latent variables, thereby simplifying the optimization by decoupling the high-dimensional inference problem into a set of lower-dimensional ones. However, it also neglects dependencies among correlated variables, potentially leading to biased approximations \citep{blei2017variational}. Specifically, we define the structured variational distribution for the parameter space $\theta$ as
\vspace{-1em}
\begin{equation}\label{eq:factorize}
q(\theta) = \{q(u_{ijl}^k, \delta_{ij}^k) , q(\omega_{ii}^k) , q(\Lambda_{ij})\}_{1 \leq i \neq j \leq p, 1 \leq k \leq K, 1 \leq l \leq L},
\vspace{-1em}
\end{equation}
which allows for joint estimation of effect sizes $u_{ijl}^k$ and their corresponding selection indicators $\delta_{ij}^k$. This formulation improves the approximation of posterior dependencies and provides uncertainty quantification, as it preserves the intrinsic coupling between the magnitude parameters and the corresponding selection indicators \citep{velten2021adaptive}.


\noindent \textbf{Augmented Transformation:} Since we place a probit prior on the selection indicators, i.e., $\delta_{ij}^k \sim \text{Bernoulli}(\Phi(\lambda_{ij}^k))$, the resulting model is non-conjugate. Conjugacy allows for analytical computation of variational expectations, thereby facilitating efficient and deterministic optimization of the evidence lower bound via coordinate-wise updates \citep{blei2017variational}. In the absence of conjugacy, these expectations often do not admit closed-form solutions. To address the intractability introduced by the probit prior, we use a data augmented framework for probit regression \citep{albert1993bayesian}. We introduce auxiliary latent variables $z_{ij}^k$ such that,  
\vspace{-1.5em}
\begin{equation*} \label{eq:augment}\vspace{-1em}
\begin{split}
 &z_{ij}^k = \lambda_{ij}^k + e_{ij}^k, \hspace{1em}
 e_{ij}^k \sim \mathcal{N}(0, 1), \\
&\delta_{ij}^k = \mathbb{I}(z_{ij}^k > 0).
\end{split}
\end{equation*}
This augmentation transforms the non-conjugate probit model into a conditionally conjugate form, restoring tractability in variational updates. The binary selection indicators $\delta_{ij}^k$ can be dealt with the continuous latent variables $z_{ij}^k$, which admit Gaussian conditional distributions. This reformulation enables efficient inference and retains compatibility with coordinate ascent updates within the structured MFVB framework (see Section \ref{suppsec:algorithm}).

\noindent \textbf{Symmetry and positive definiteness}: To ensure the symmetry of the estimated conditional dependence matrices, i.e., $\omega_{ij}(s^k) = \omega_{ji}(s^k)$ for each slide $k$, we adopt a post-processing strategy inspired by the symmetrization rules in \citep{meinshausen2006high}. Let the posterior inclusion probability (PIP) for the edge between nodes $i$ and $j$ in FOV $k$ be defined as $\hat{p}_{ij}^k = \hat{p}_{ji}^k = \min\left\{\mathbb{E}(\delta_{ij}^k), \mathbb{E}(\delta_{ji}^k)\right\}$. The conditional dependency exists, i.e., $\omega_{ij}(s^k) \neq 0$, if $\hat{p}_{ij}^k$ exceeds a threshold $\kappa_\alpha$. To determine $\kappa_\alpha$, we use a Bayesian local false discovery rate (FDR) control procedure that aims to control the average Bayesian FDR at a pre-specified level $\alpha$ \citep{baladandayuthapani2010bayesian}. For edges that satisfy $\hat{p}_{ij}^k > \kappa_\alpha$, we define 
$\tilde{u}_{ij}^k = \mathbb{E}(\omega_{ii}^k) \cdot \mathbb{E}(u_{ij}^k)$. To enforce symmetry, we define the final symmetrized estimate by selecting the one with smaller $l_1$ norm $\hat{u}_{ij}^{k} = \hat{u}_{ji}^{k} = \tilde{u}_{ij}^k \cdot \mathbb{I}\left(|\tilde{u}_{ij}^k|_1 < |\tilde{u}_{ji}^k|_1\right) + \tilde{u}_{ji}^k \cdot \mathbb{I}\left(|\tilde{u}_{ij}^k|_1 \geq |\tilde{u}_{ji}^k|_1\right).$

As the off-diagonal entries in the precision matrix are spatially varying, a natural sufficient condition for ensuring positive definiteness of $\Omega(s)$ is to make it diagonally dominant, $\forall s\in S$. We implement a rescaling step in post processing procedure of $\hat{u}_{ij}^k$ \citep{zhang2022high} to ensure 
$
    \sum_{j \neq i}^{p} \left| \sum_{l=1}^{L} \hat{u}_{ijl}^{k} B_l(s^k) \right| \leq \omega_{ii}^k.
$
Let $B^k = [B_{1}(\mathbb{S}^k), \dots, B_{L}(\mathbb{S}^k)]$. Define $a^k = \|B^k\|_{\infty} < \infty $, where $\|\cdot\|_{\infty}$ represents the sup-norm. Since $B^k$ represents the basis functions in the Gaussian process described in Section~\ref{suppsec:method_gp}, the quantity $a^k$ is an observed value that depends on spatial location and the Gaussian process parameters. 
We then rescale the coefficients $\hat{u}_{ijl}^{k}$ by the factor $a_{k} \sum_{j \neq i,\, 1 \leq l \leq L} |u_{ijl}^{k}|$ to satisfy the above condition. A detailed proof is provided in the Section \ref{suppsec:msgr_post_steps}.


\vspace{-2em}
\section{Simulation Studies}\label{sec:simStudy}
\vspace{-1em}
To demonstrate the effectiveness of our proposed \texttt{mSGR} method, we conduct replicated simulations studies under various simulation settings to evaluate the performance in terms of graph structural recovery. We begin by outlining the data generation process, which mimics the multi-resolution architecture of spatial transcriptomics data, followed by descriptions of the competing methods and structural recovery measures. We assess the performance of \texttt{mSGR} under two scenarios: a synthetic setting under varying sample sizes and spatial correlations (Scenario I) and a sRCC data-informed simulation incorporating spatially varying edges (Scenario II). All results are aggregated over 50 independent replicates.

\vspace{-1.25em}
\paragraph{Data generation for scenario I:} \hspace{-1em}
We fix the number of genes at $p = 30$ with sparsity at $5\%$ and vary the sample size for each FOV $N_k \in \{500, 1000, 1500\}$. The spatial correlation matrix $\tilde{V} \in \mathbb{R}^{K \times K}$ is defined with entries $\tilde{V}[k,k'] = \rho^{d_{kk'}}$, where $d_{kk'}$ denotes the Euclidean distance between centroids of FOVs $k$ and $k'$ and spatial correlation are ranged from $\rho \in \{0.3, 0.6, 0.9\}$. FOVs are arranged in a \( 5 \times 5 \) grid, each occupying a \( 0.4 \times 0.4 \) unit square, with 0.6-unit spacing between adjacent regions. Within each FOV \( k \), we generate a \( 40 \times 40 \) regular grid of candidate cell positions and randomly sample \( N_k \) cells without replacement, yielding spatial coordinates \( \mathbb{S}^k = \{s_{n_k}^k\}_{n_k=1}^{N_k} \subseteq \mathbb{R}^2 \). For each gene pair $(i,j)$, $\Lambda_{ij} \in \mathbb{R}^K$ is generated from a multivariate normal distribution i.e. $\Lambda_{ij} \sim \mathcal{N}(0, \tilde{V})$, based on which selection indicators are defined using a threshold to control the sparsity. For each included edge, we consider a spatial function to quantify the off-diagonal entries and diagonal entries are adjusted to ensure positive definite property of precision matrices. The gene expression vectors are generated from the resulting multivariate Gaussian distributions.

\vspace{-1.25em}
\paragraph{Data generation for scenario II:} \hspace{-1em} 
We adopt a similar structure from scenario I while mimicking real data based settings. The spatial coordinates and the number of cells within each FOV are considered from the RCC dataset in Section \ref{sec:Application}. We fix the number of genes at $p = 27$ and set the sparsity level to 12\%, consistent with our observation from RCC data. To construct spatially varying edges, we construct a library of smooth functions derived from the estimated spatial effects $\{\hat{\gamma}_{ij}\}$ in RCC dataset. For each function, we compute its maximum absolute value across cells within FOVs and retain those with maximum magnitude exceeding $0.5$, and scale them by a factor of $2$. Each selected edge is randomly assigned to one of these rescaled functions to define the spatially varying off-diagonal elements in the precision matrix. To guarantee positive definiteness, we iteratively add a small quantity of $0.2$ to all diagonal entries until the resulting matrix becomes positive definite. Finally, we simulate the gene expression vectors from multivariate Gaussian distributions with derived spatially varying precision matrices.

\vspace{-1.25em}
\paragraph{Comparison Methods:} \hspace{-1em} 
We compare our proposed method against several state-of-the-art approaches for estimating multiple graphical models. Spatial Graphical Regression (SGR, \cite{acharyya2025spatially}) is a spatially aware method to estimate spatially varying graphs but it is limited to a single contiguous region and does not account for multiple FOVs. GraphR \citep{chen2025probabilistic} is a probabilistic graphical regression framework that incorporates external covariates through a linear design matrix. 
We also compare \texttt{mSGR} with existing mGGM-based approaches. 
Joint Graphical Lasso (JGL, \cite{danaher2014joint}) and jointGHS \citep{lingjaerde2024scalable} are multi-group Gaussian graphical models that estimate precision matrices for each FOV jointly with a group-fused penalty or hierarchical shrinkage, respectively. However, they assume discrete groups and do not incorporate spatial distances across regions. Finally, we include a baseline graphical lasso (Glasso, \cite{friedman2008sparse}) method applied independently to each FOV, which does not borrow information across FOVs and ignores spatial context entirely. 
\vspace{-1.25em}
\paragraph{Evaluation Metrics:} \hspace{-1em} For comprehensive assessment of graph structural recovery, we use three performance metrics: Matthews correlation coefficient (MCC), true positive rate (TPR), and false discovery rate (FDR). MCC incorporates all four components of confusion matrix components—providing a summary statistic reflecting structural recovery performance. Ranging from -1 to 1, an MCC of 1 indicates perfect classification, and -1 reflects total disagreement. In the context of sparse graph recovery, MCC is a robust metric for evaluating binary classification performance, especially in imbalanced settings \citep{matthews1975comparison, chicco2020advantages}.


\begin{figure}[ht!]
  \centering
  \includegraphics[width=\textwidth]{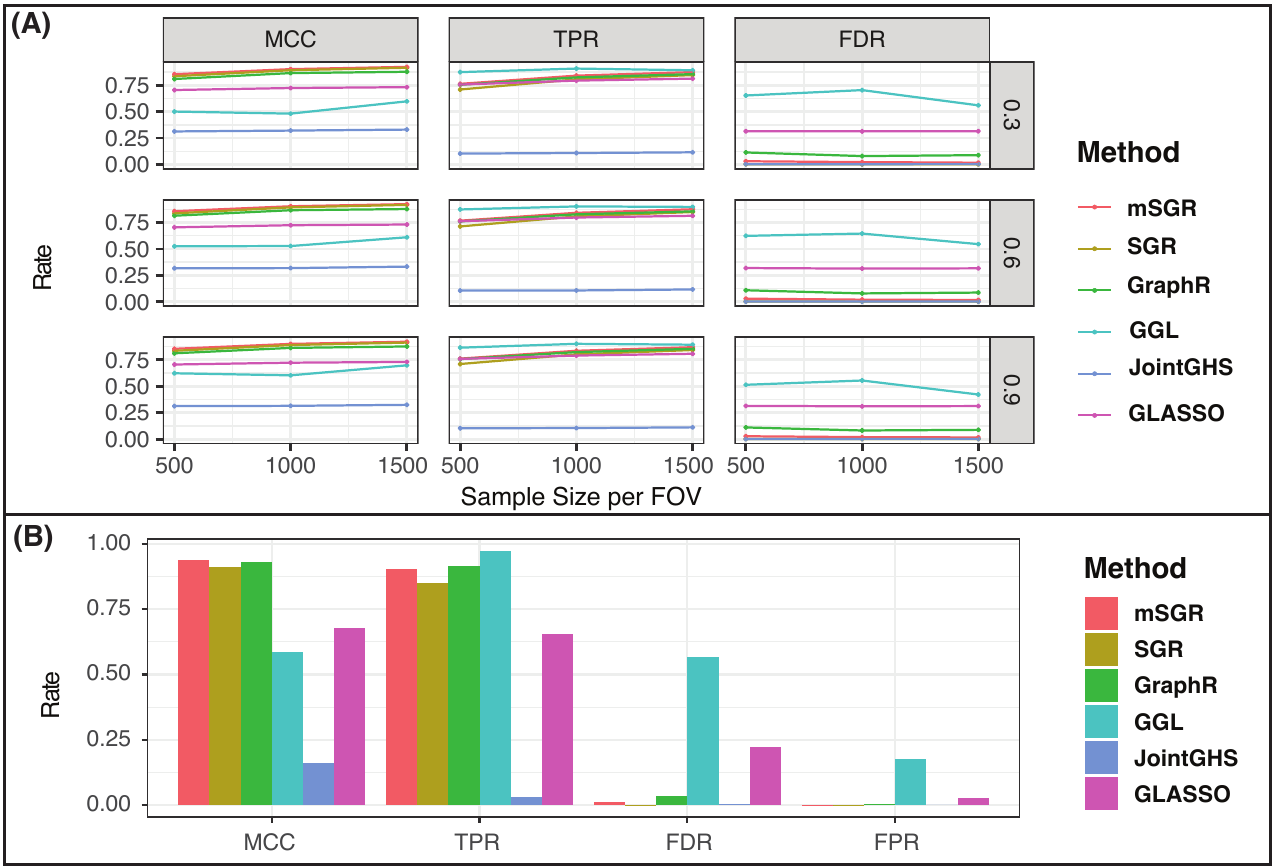}
  \caption{\small \textbf{Simulation performance of \texttt{mSGR}.} 
\textbf{(A)} Simulation results comparing different methods under varying correlation strengths between FOVs ($\rho$ $=$ $0.3$, $0.6$, $0.9$) and sample sizes ($n=500,1000,1500$ per FOV). Figure \ref{fig:simulation-performance}\textbf{(B)} shows competitive performance of \texttt{mSGR} for graph structral recovery in the real data–based simulation. The plot represents mean values of the  evaluation metrics (MCC, TPR, FDR, FPR) across replications.}
  \label{fig:simulation-performance}
  \vspace{-1em}
\end{figure}

\vspace{-1.25em}
\paragraph{Simulation results of scenario I:} \hspace{-1em} Figure \ref{fig:simulation-performance}A presents the simulation results under various degrees of correlation between FOVs and sample sizes. At a moderate correlation level ($\rho$ = 0.6) and a sample size of 1000, \texttt{mSGR} achieves the highest value of MCC (\texttt{mSGR}: $0.9$, SGR: $0.89$, GraphR: 0.87, GLASSO: 0.72, GGL: 0.53, JointGHS: 0.32) among other competing methods. The \texttt{mSGR} method achieves moderately high TPR (\texttt{mSGR}: $0.84$, SGR: $0.81$, GraphR: $0.83$, GLASSO: $0.8$, GGL: $0.9$, JointGHS: $0.11$) while balancing for low FDR (\texttt{mSGR}: $0.021$, SGR: $0.002$, GraphR: $0.08$, GLASSO: $0.314$, GGL: $0.645$, JointGHS: $0.002$). GGL shows high TPR although it suffers from the highest FDR of 0.65 among all methods. SGR yields marginally lower FDR than \texttt{mSGR} though it comes at the cost of lower TPRs. \texttt{mSGR} achieves lower FDR than all other methods. In summary, \texttt{mSGR} consistently demonstrates balanced performance in terms of high precision, and reasonable control over false positives, outperforming other methods in MCC. 

\vspace{-1.25em}
\paragraph{Simulation results of scenario II:} \hspace{-1em}
Figure \ref{fig:simulation-performance}B demonstrates the performance of \texttt{mSGR} under real data based simulation settings. The proposed \texttt{mSGR} (MCC: $0.94$, TPR: $0.90$, FPR: $0.001$, FDR: $0.01$) attains the highest MCC, reflecting a high level of agreement with the true underlying spatially varying graphs, along with a high TPR and very low error rates. The \texttt{GraphR} (MCC: $0.93$, TPR: $0.91$, FPR: $0.004$, FDR: $0.03$) method achieves lower MCC and marginally higher sensitivity, although at the cost of reduced precision, as indicated by higher levels of FDR and FPR. \texttt{SGR} (MCC: $0.91$, TPR: $0.85$, FPR $<$ $0.001$, FDR: $0.001$) yields competitive results, with lower MCC than \texttt{mSGR} and very low FDR, FPR, but it shows relatively lower detection power with TPR. In contrast, the remaining methods exhibit performance limitations. GGL (MCC: $0.59$, TPR: $0.97$, FPR: $0.17$, FDR: $0.57$) achieves a high TPR, but its precision suffers significantly, as reflected through low FDR and MCC. GLASSO (MCC: $0.68$, TPR: $0.66$, FDR: $0.22$, FPR: $0.03$) yields a comparatively similar performance. The method jointGHS (MCC: $0.16$, TPR: $0.03$, FDR: $0.002$, FPR: $\approx 0$) provides very limited performance with overly conservative edge selection that fails to recover meaningful network structures despite minimal error rates.


\vspace{-1.25em}
\paragraph{Computation time and additional simulation studies:} \hspace{-1em}
In a high-performance computing environment, the average runtime per replicate for \texttt{mSGR} in Simulation~I is approximately 0.34 hours with 500 samples per FOV, 1.26 hours with 1{,}000 samples per FOV, and 2.43 hours with 1{,}500 samples per FOV. In Simulation~II, the average runtime per replicate is approximately 1.34 hours. For reference, running the application on a local machine (MacBook Pro equipped with an Apple M4 Pro chip) requires approximately 7 minutes. Further computational details are provided in the Section \ref{suppsec:computation_time}.  We performed a detailed sensitivity analyses to assess the robustness of \texttt{mSGR} to prior specification and model flexibility. For different settings of Gaussian process hyperparameters and sparsity levels, \texttt{mSGR} shows consistent trends in MCC, TPR, FDR, and FPR as reported in the Section \ref{suppsec:additional_sim1}.

\vspace{-2em}
\section{Multi-resolution spatial network characterization in renal cell carcinoma}
\label{sec:Application}
\vspace{-1em}
\paragraph{Scientific background:} \hspace{-1em} Renal cell carcinoma (RCC) is the most common type of kidney cancer \citep{cirillo2024global}. In 2023, an estimated 81,800 new cases were diagnosed in the United States, making it the sixth most common cancer among males and ninth most common among females. Globally, RCC is the 15th leading cause of cancer-related death, with over 179,000 deaths reported in 2020 \citep{rose2024renal}. Sarcomatoid renal cell carcinoma (sRCC) is a de-differentiation of primary RCC which arises through epithelial-mesenchymal transition (EMT) \citep{vcugura2024epithelial}. While sRCC represents the most extreme version of EMT in RCC, EMT plays a critical role in all RCC tumors broadly and is associated with an increased risk \citep{piva2016epithelial}. 
However, EMT is not a uniform process; rather, it occurs in concert with dynamic changes in the TME \citep{brabletz2021dynamic}. A key scientific question, therefore, is how EMT and its interactions with the TME are organized within tumor architecture. While EMT in RCC can be highly heterogeneous, sRCC represents an ideal model to study EMT gradients, as it exhibits a structured progression of EMT that can be detected histologically and interrogated at the microscale. Therefore, spatially characterizing EMT within the TME provides a unique opportunity to uncover tumor gradients that are obscured in bulk analyses, offering deeper insights into mechanisms of tumor aggressiveness and potential therapeutic vulnerabilities.


\vspace{-1.25em}
\paragraph{Multi-resolution ST data:} \hspace{-1em} Our case study arises from high-resolution hierarchical ST data from RCC tissue using the CosMx platform \citep{may2025spatial}. The dataset includes 23 FOVs across 960 genes, capturing a total of 26,298 cancer cells, with cell counts per FOV ranging from approximately 460 to 1,800. These FOVs are categorized into three histopathological subtypes: 8 sarcomatoid, 12 clear cell, and 3 transitional regions exhibiting intermediate phenotypic characteristics between clear cell and sarcomatoid types. 
We investigate multiple biological pathways, with a primary focus on the EMT pathway ($p=27$ genes), which plays a critical role in RCC progression, invasion, and therapeutic response. For our investigation, we focus on cancer cells to delineate how gene regulatory patterns change across the spatial EMT gradient. Further details of the data preprocessing steps and choice of model hyperparameters can be found in Supplementary Section \ref{suppsec:application}. 






\begin{figure}[ht!]
  \centering
  \includegraphics[scale=0.3]{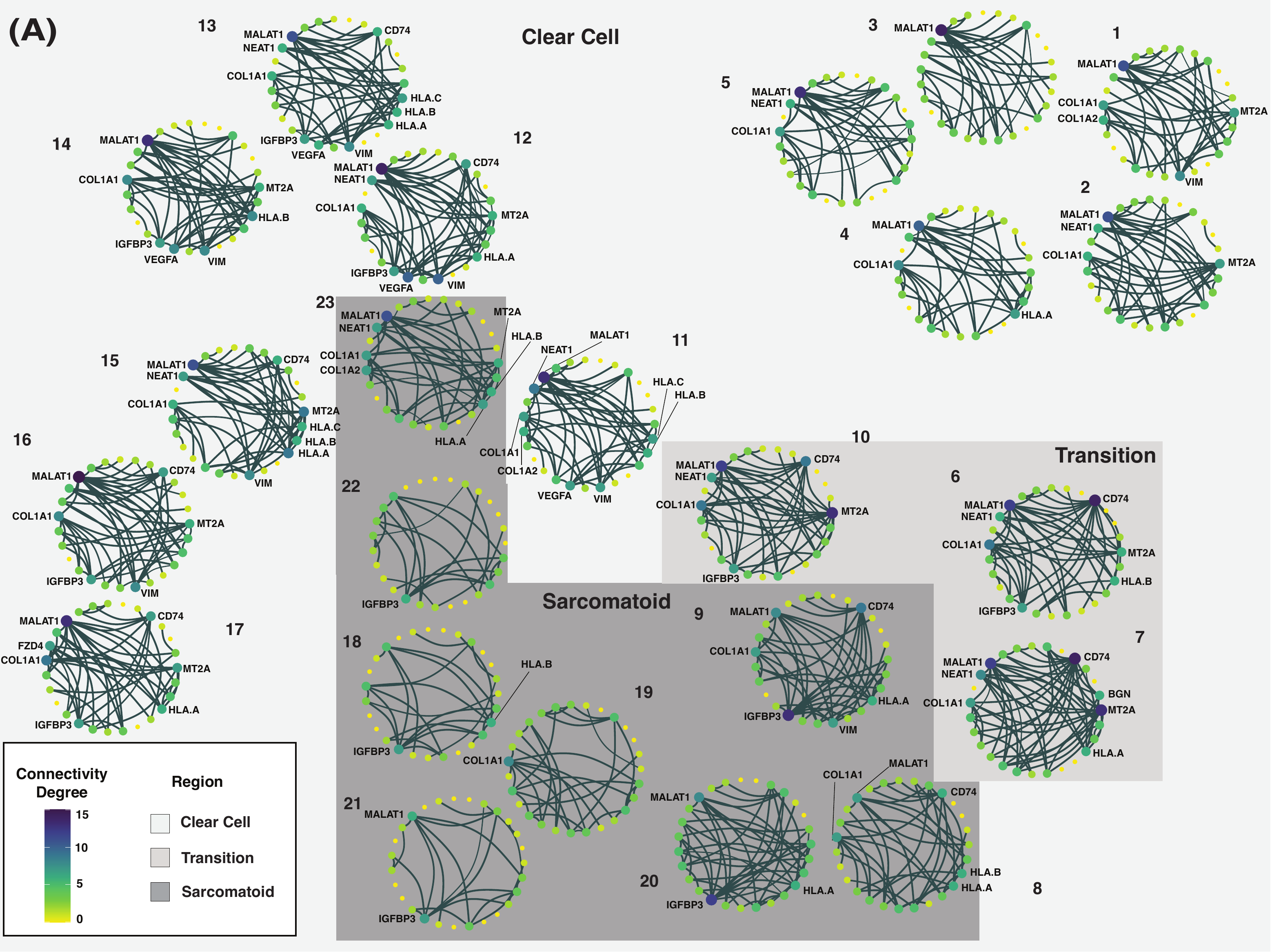}
  \vspace{-0.5em}
  \caption{\small \textbf{EMT Gene Networks Across Various Regions in RCC.} \textbf{(A)} Networks depict conditional dependencies among genes from EMT pathway in clear cell, transitional, and sarcomatoid RCC regions. Nodes represent genes, and edges indicate conditional associations inferred from multi-resolution ST data. Node color corresponds to connectivity degree which is defined as the summation of significant connection, while posterior inclusion probabilities (PIP) quantify edge width. Gene names are displayed only for nodes with connectivity degree greater than 5. Each network is annotated with its corresponding FOV-specific number.}
  \label{fig:emt_networks_all}
  \vspace{-1em}
\end{figure}


\vspace{-1.25em}
\paragraph{Spatially varying graphs across multiple FOVs:} \hspace{-1em} We apply the \texttt{mSGR}  on RCC data, and Figure \ref{fig:emt_networks_all} shows spatially varying graphs of EMT pathway genes across different spatial domains of  clear cell, transitional, and sarcomatoid regions (marked in different background colors). We observe particularly dense networks in the transitional region, as well as in clear cell, and sarcomatoid areas located near the transition zone (FOVs 8, 9, 20, and 23). 
While marginal correlations among mesenchymal genes are generally high in sarcomatoid regions 
due to the global upregulation of EMT programs \citep{bostrom2012sarcomatoid}, their conditional dependence structure is relatively weaker, as much of the co-expression appears to arise from shared associations with largely mediated through shared links with extracellular matrix remodeling and immune-interaction genes—such as \textit{COL1A1}, rather than direct regulatory interactions \citep{nieto2016emt}. 

\begin{figure}[ht!]
  \centering
  \includegraphics[scale=0.7]{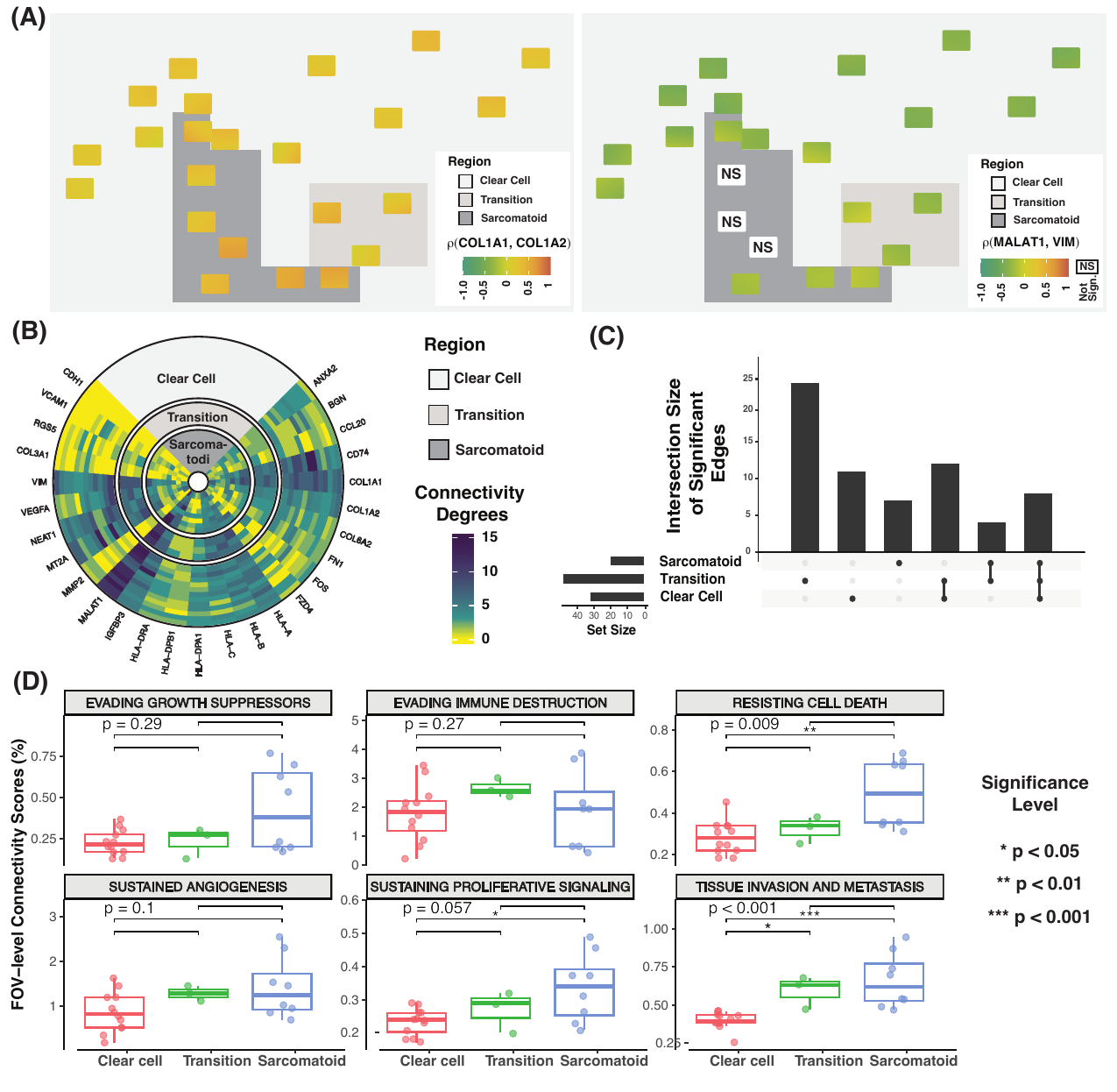}
  \vspace{-0.5em}
  \caption{\small \textbf{Genomic network-based characterization of region in RCC} (A) Spatial patterns of partial correlations for selected gene pairs (COL1A1–COL1A2; MALAT1–VIM), with background color distinguishing clear cell, transitional, and sarcomatoid regions. (B) Circular heatmap of connectivity degrees across genes within each region, where connectivity degree is defined as the total number of significant edges incident to a node. (C) UpSet plot of significant connections across regions, summarizing the number of significant gene–gene connections detected in each region and their overlaps. A connection is considered significant in a region if it is significant in more than 50\% of the FOVs belonging to that region.
  (D) Boxplots showing the distribution of field-of-view (FOV)–level connectivity degree across spatial regions. The connectivity scores (CS) is defined as $\mathrm{CS} = \#~\text{significant edges}/\#~\text{possible edges} \times 100$. Each point represents one FOV, with regions color-coded. 
    The overall difference among regions within each pathway facet was assessed using the Kruskal--Wallis test (p-values shown at the top), 
    and pairwise differences were evaluated by Wilcoxon rank-sum tests with Benjamini--Hochberg adjustment. 
    Significance levels are indicated by stars (* $p < 0.05$, ** $p < 0.01$, *** $p < 0.001$).}
  \label{fig:emt_others}
  \vspace{-1em}
\end{figure}

\vspace{-1.25em}
\paragraph{Spatially varying edges:} \hspace{-1em} We present the spatially varying partial correlations of selected gene pairs in Figure \ref{fig:emt_others}A, demonstrating the ability of \texttt{mSGR} to incorporate spatial information and explicitly estimate heterogeneous dependence structures across regions. Notably, we observe a consistently high partial correlation between COL1A1 and COL1A2 across all regions, with the strongest signal emerging in the sarcomatoid area. This finding is biologically consistent, as the two genes encode the $\alpha 1$ and $\alpha 2$ chains of type I collagen, which physically assemble into a heterotrimeric triple helix \citep{li2016identification}. The statistical evidence of direct conditional dependence is concordant with their well-established structural interdependence in forming type I collagen. In the context of sarcomatoid, this tight coupling likely reflects enhanced extracellular matrix remodeling, which is closely associated with tumor progression and metastasis \citep{walker2018role, eble2019extracellular}. 

\vspace{-1.25em}
\paragraph{Novel spatial biomarker identification:} \hspace{-1em} We also identify key hub genes by quantifying connectivity degrees, defined as the total number of significant edges connected to each node. Figure \ref{fig:emt_others}B highlights that MALAT1 and VIM consistently act as hub genes across all regions, exhibiting higher connectivity degrees in clear cell compared to sarcomatoid regions (also see Supplementary Figure \ref{suppfig:hub_connectivity}). Additionally, to identify region-dependent hub patterns, we perform a nonparametric test to compare the connectivity degree across clear cell, transition, and sarcomatoid regions for all genes. Five genes (HLA-DPA1, MALAT1, MT2A, NEAT1 and VCAM1) show significant regional differences after multiple testing correction, indicating distinct patterns along the EMT gradient. Boxplots illustrating these differences are provided in Supplementary Figure \ref{suppfig:connecticity_difference}.

Notably, MALAT1 and VIM show a strong negative partial correlation in clear cell regions, which becomes weaker in transition and sarcomatoid regions (Figure \ref{fig:emt_others}A right). MALAT1 is frequently upregulated in clear cell RCC and has been linked to poor prognosis \citep{zhang2015upregulation}, although its role in sarcomatoid RCC remains less well characterized. In our data, MALAT1 expression is enriched in clear cell and transitional regions but decreases in sarcomatoid regions (Supplementary Figure \ref{suppfig:expression_difference}) suggesting that its prognostic association may not directly translate to sarcomatoid differentiation. By contrast, VIM, a canonical mesenchymal marker—shows moderate expression in clear cell regions but strong enrichment in sarcomatoid regions \citep{vcugura2024epithelial}, consistent with its established role in mesenchymal activation. The reciprocal expression and negative conditional association between MALAT1 and VIM thus highlight an antagonistic relationship that evolves spatially along the tumor gradient, reflecting the dynamic balance between epithelial maintenance and mesenchymal activation during tumor progression. 

Transitional regions exhibit stronger and more specific partial correlations, as illustrated in Figure \ref{fig:emt_others}C, where 47 significant edges (23 region-specific) are detected, compared with 31 (11 region-specific) in clear cell and 19 (7 region-specific) in sarcomatoid regions. Given that sarcomatoid differentiation has been strongly linked to EMT activation in RCC \citep{conant2011sarcomatoid}, the denser and more specific connectivity observed in transitional regions likely reflects dynamic regulatory rewiring during EMT progression, capturing the acquisition of mesenchymal features \citep{li2023heterogeneity, pal2021partial}. These findings demonstrate that our framework captures spatial variation in network topology and identifies region-specific key regulators of RCC progression.

\vspace{-1.25em}
\paragraph{Connectivity between genes in hallmark pathways:} \hspace{-1em} 
We next quantify the pathway coordination of 6 canonical hallmark pathways of cancer progression \citep{Hanahan2022hallmarks}.  We use connectivity score (CS), defined as the proportion of significant partial correlations among all possible gene–gene pairs within a pathway. This metric captures the extant of coordinated molecular activity and provides a pathway-level summary of regulatory tightness \citep{chen2025probabilistic}. Across regions, CS values reveal distinct functional remodeling along the clear cell–to–sarcomatoid transition (Figure \ref{fig:emt_others}D). These differences are statistically significant for both the resisting cell death ($p = 0.009$) and the tissue invasion and metastasis pathways ($p < 0.001$). For the resisting cell death pathway, the median CS increases from 0.28\% (range: 0.18\%-0.45\%) in clear cell regions to 0.34\% (range: 0.25\%-0.38\%) in transition regions and 0.5\% (range: 0.31\%-0.69\%) in sarcomatoid regions. A similar pattern is observed for the tissue invasion and metastasis pathway, where sarcomatoid regions display consistently elevated connectivity (median 0.62\%; range: 0.47\%–0.95\%) relative to clear cell regions (median 0.4\%; range: 0.25\%–0.46\%). These observations align strongly with the known aggressive biology of sRCC, which is characterized by increased motility, invasive capacity, and a high resistance to apoptosis \citep{lebacle2019epidemiology}. Moreover, sarcomatoid regions display a wider range of CS values across FOVs for most hallmark pathways, indicating greater heterogeneity in pathway-level network organization, consistent with previous findings that the mesenchymal component of sRCC exhibits a highly variable appearance \citep{lebacle2019epidemiology}.


\vspace{-2em}
\section{Discussion}\label{sec:Discussion}
\vspace{-1em}
In this paper, we introduce the \texttt{mSGR} framework, a hierarchical Bayesian model for spatial network estimation of high-dimensional, and multi-resolution ST data. \texttt{mSGR} jointly infers multiple spatially varying graphs while borrowing information across spatial resolutions to capture both local heterogeneity and global spatial network structures. 
This is achieved through a novel Bayesian structured edge selection method, which regulates the inclusion probabilities of edges across spatial surfaces, allowing the graph topology to borrow strength from neighboring regions according both spatial distance and region annotations. The GP prior on spatially varying regression coefficients enables partial correlations to vary smoothly across spatial regions. This dual mechanism provides a coherent way to integrate global-local heterogeneity and multi-resolution spatial graphical inference. 

Methodologically, \texttt{mSGR} generalizes existing spatial and multiple-GGM frameworks by embedding spatially structured Bayesian edge selection into the graphical model, and
enabling spatially coherent yet locally adaptive edge selection. 
To achieve scalable and tractable inference, \texttt{mSGR} employs an MFVB algorithm built upon an augmentation representation that converts non-conjugate selection indicator priors into conditionally conjugate forms. This reparameterization allows closed-form variational updates, and enables computational efficiency while maintaining inference accuracy. Together, these innovations yield a unified, interpretable, and computationally efficient framework for analyzing spatially heterogeneous gene networks from multi-resolution designs.

 %

Although we focus on undirected multiple spatially varying graphs, the \texttt{mSGR} framework can be adapted to more general classes of directed graphs with a known genomic ordering. Notably, this framework naturally extends to other continuous covariates, such as pathological gradients or clinical measures.  From an inferential perspective, \texttt{mSGR} offers a principled way to quantify uncertainty in multiple spatially resolved network structures and to assess region-level connectivity metrics such as connectivity degrees and pathway-specific summaries. The hierarchical formulation also naturally accommodates multi-region comparisons and facilitates formal testing of region-dependent network alterations. While the motivating application focuses on ST data in cancer, the \texttt{mSGR} framework is broadly applicable to other domains involving spatially structured multivariate data, such as neuroimaging and environmental monitoring. Notably, \texttt{mSGR} also supports characterization of spatial networks across multiple (tissue) samples; in such settings, careful calibration of the sample-correlation prior \(V\) is important for robust borrowing of information across samples. Finally, \texttt{mSGR} can be extended to characterize networks for other spatial omics data and across spatial multi‑omics layers; we leave these tasks for future consideration. 

\vspace{-0.5em}
\begin{center}
{\large\bf SUPPLEMENTARY MATERIALS}
\end{center}

\noindent \textbf{Title}: The Supplementary PDF document provides additional methodological, computational and implementation details for \texttt{mSGR} along with additional results of simulation studies and ST data analysis. 

\noindent \textbf{R-package}: The R package is available at \url{https://github.com/***}.

\noindent \textbf{Disclosure Statement}:
The authors report there are no competing interests to declare.

\noindent \textbf{Data Availability Statement}:
Raw data of RCC are deposited in the Gene Expression Omnibus (GEO) under accession numbers GSE299368, and will be publicly available upon publication of the corresponding manuscript. 

\noindent \textbf{Use of Generative AI:} The authors used ChatGPT (OpenAI, GPT-5.2) to assist with language editing during manuscript preparation. The tool was used solely to refine grammar and improve readability and did not contribute to the methodology, analysis, or scientific conclusions. All authors have reviewed and approved the final manuscript.

\vspace{-2em}
\bibliography{msgr.bib}

\newpage
\title{\Large \bf Supplementary for ``Multi-resolution Spatial Graphical Regression Models for Hierarchical Spatial Transcriptomics Data"}

\setcounter{table}{0}
\renewcommand{\thetable}{S\arabic{table}}

\setcounter{figure}{0}
\renewcommand{\thefigure}{S\arabic{figure}}

\setcounter{section}{0}
\renewcommand{\thesection}{S\arabic{section}}

\setcounter{equation}{0}
\renewcommand{\theequation}{S\arabic{equation}}

\noindent
This Supplementary Material provides additional details to support the methodology, computation, and empirical results. Section~\ref{suppsec:msgr_method} presents the \texttt{mSGR} model formulation and key methodological components, including the Gaussian process prior with the modified squared exponential kernel, model specification, priors and hyperparameters, initialization, the likelihood, derivation of closed-form MFVB updates, the full variational inference algorithm, and postprocessing steps such as Bayesian FDR thresholding and positive-definiteness enforcement. Section~\ref{suppsec:simulation} provides additional simulation study details, including evaluation metrics, competing methods, and sensitivity analyses. Finally, Section~\ref{suppsec:application} gives supplementary details for the real-data application on EMT gradient characterization in sarcomatoid renal cancer, including preprocessing/normalization, incorporation of spatial coordinates at both FOV and cell levels, hyperparameter choices, and additional figures.

\section{Methodological details for \texttt{mSGR}} \label{suppsec:msgr_method}

\subsection{Gaussian process prior with modified exponential sqaured covariance kernel} \label{suppsec:method_gp}

We utilize Gaussian processes (GPs) to model spatial correlations on micro (cell) level.  
GPs serve as a fundamental tool for modeling complex spatial dependencies \citep{cressie2011statistics, banerjee2003hierarchical}. In this context, we consider a zero-mean GP prior defined over a spatial domain $\mathbb{S}^k \subseteq \mathbb{R}^2$, characterized by a covariance kernel $\mathbf{C}(\cdot, \cdot): \mathbb{R}^{2 \times2} \rightarrow \mathbb{R}$, such that $\beta(\cdot) \sim \mathcal{GP}(\mathbf{0}, \sigma^2 \mathbf{C}(\cdot, \cdot))$. Under this formulation, any finite collection of spatial locations $s^k, t^k \in \mathbb{S}^k$ yields random variables that are jointly Gaussian distributed with covariance given by $\sigma^2\mathbf{C}(s^k, t^k)$.

In this work, we adopt a GP prior with a Modified Squared Exponential (MSE) kernel \citep{shi2015thresholded, lin2024latent} defined as:
\begin{align}
    \label{eq:gp}
&\beta_{ij} (\cdot) \sim \mathcal{GP}(\mathbf{0},\sigma^2_{ij}\mathbf{C}(\cdot, \cdot) \nonumber
\\
    &\mathbf{C}(s^k, t^k) = \exp\left\{-a_{gp}(||s^k||_2^2 + ||t^k||_2^2) - b_{gp} ||s^k - t^k||_2^2 \right\}, \nonumber \\ &\text{ for prespecified } a_{gp}, b_{gp} >0 \text{, and } s^k, t^k\in \mathbb{S}^k.
\end{align}

Here $\|\cdot\|_2$ denotes the Euclidean norm. This kernel generalizes the classical squared exponential kernel by introducing an additional exponential decay factor dependent on the norms of $s^k$ and $t^k$. From the specified Gaussian process, the following properties are easily derived:
\begin{align}
    &\text{Var}[\beta_{ij}(0)] = \sigma^2_{ij} \nonumber \\
    &\text{Var}[\beta_{ij}(s^k)] = \sigma^2_{ij} \text{ exp}(-a_{gp} \| s^k\|_2^2)\nonumber \\
    &\text{Cor}[\beta_{ij}(s^k),\beta_{ij}(t^k)] = \text{ exp}(-b_{gp} \| s^k - t^k\|_2^2) 
\end{align}

Specifically, $\sigma_{ij}^2$ controls the maximum marginal variance of the process, setting the overall scale of variability for $\beta_{ij}(\cdot)$. The term exp$(-a_{gp}(||s^k||_2^2 + ||t^k||_2^2)$ imposes global attenuation that diminishes influence far from the origin, with larger $a_{gp}$ corresponds to a faster decay rate of $\text{Var}[\beta_{ij}(s^k)]$ to $\text{Var}[\beta_{ij}(0)]$. Notably, when $a_{gp} =0$, the modified squared exponential kernel becomes a standard exponential kernel. Meanwhile, the term exp$(-b_{gp}||s^k - t^k||_2^2)$ captures local smoothness and spatial proximity, with a larger $b_{gp}$ corresponding to a smoother $\beta_{ij}(s^k)$, and thus impose stronger correlation between nearby points. 

The primary barrier to implementing GPs lies in their significant computational and numerical challenges. For each field of view (FOV), GP inference requires inversion of an  $N_k \times N_k$ covariance matrix, incurring a computational complexity of $O(N_k^3)$, which becomes prohibitively expensive as the dataset size grows. In addition to the computational cost, the inversion of the covariance matrix is numerically unstable, particularly when the matrix is ill-conditioned due to near-collinearity in the inputs or unfavorable kernel parameter settings \citep{banerjee2013efficient}. To facilitate computation and representation, we utilize Mercer’s theorem \citep{minh2006mercer, shi2015thresholded}, which states that under mild regularity conditions, a continuous, symmetric, and positive semi-definite kernel admits a spectral decomposition in terms of its eigenfunctions and eigenvalues. This result enables the application of the Karhunen–Loève (KL) expansion \citep{wang2008karhunen}, expressed as
\begin{align}
    \beta_{ij}(s^k) = \sum_{l=1}^ \infty u^k_{ijl} \sqrt{\eta_l} \psi_{l}(s^k) := \sum_{l=1}^ \infty u^k_{ijl} B_l(s^k)\label{sup_eq:kl}
\end{align}
where $\eta_l$ and $\psi_{l}(\cdot)$ are the eigenvalue-eigenfunction pairs of the covariance kernel, and $u^k_{ijl} \overset{\text{i.i.d.}}{\sim} \mathcal{N}(0, \sigma_{ij}^2)$. This representation provides a compact and flexible approximation of the underlying stochastic process. In practical implementations, the infinite series is truncated to the first $L$ dominant components, thereby reducing the dimensionality from $N_k$ to $L$ and enabling scalable inference in high-dimensional spatial models.

\subsection{Augmented Model for Selection Indicators}
As shown in Equation \ref{eq:augment}, we introduce a latent variable $z_{ij}^k = \lambda_{ij}^k + \epsilon_{ij}^k$, where $\epsilon_{ij}^k \sim N(0,1)$. The selection indicator is then expressed as $\delta_{ij}^k = I(z_{ij}^k > 0)$. Consequently,
\newline
$P(\delta_{ijk} = 1) = P(z_{ij}^k > 0) = P(\lambda_{ij}^k + \epsilon_{ijk} > 0)  = P(\epsilon_{ijk} > -\lambda_{ij}^k) = \Phi(\lambda_{ij}^k)$, where $\Phi(\cdot)$ denotes the cumulative distribution function of the standard normal distribution.

\subsection{Mean Field Variation Bayes}
Suppose $\theta = \left\{u_{ijl}^k, \delta_{ij}^k, \omega_{ii}^k, \Lambda_{ij} \right \}_{1 \leq i \neq j \leq p, 1 \leq k \leq K, 1 \leq l \leq L}$ denote the parameters of interest. The true posterior distribution is $p(\theta \mid \mathcal{D})$, where $\mathcal{D}$ is the observed data. The goal is to approximate $p(\theta \mid \mathcal{D})$ with a structured variational distribution $q(\theta) \in \mathcal{Q}$, where $\mathcal{Q}$ includes factorizations that retain dependencies within predefined variable blocks, e.g., 
\begin{align}\label{eq:factorize}
    q(\theta) = q(u_{ijl}^k, \delta_{ij}^k) q(\omega_{ii}^k)q(\Lambda_{ij})
\end{align}
The optimal variational distribution $q^\ast(\theta)$ minimizes the KL divergence to the true posterior:
\begin{align*}\
q^\ast(\theta) = \arg\min_{q \in \mathcal{Q}} \mathrm{KL}\left(q(\theta) \,\|\, p(\theta \mid \mathcal{D})\right).
\end{align*}
This is equivalent to maximizing the evidence lower bound (ELBO):
\begin{align*}\
\mathcal{L}(q) = \mathbb{E}_{q(\theta)}[\log p(\mathcal{D}, \theta)] - \mathbb{E}_{q(\theta)}[\log q(\theta)]
\end{align*}
\newline
To improve numerical stability, all variational updates are computed using the full dataset. At iteration $t$, each variational parameter is updated according to
\begin{equation}
\boldsymbol{\theta}^{(t)}
=
\rho\, \boldsymbol{\theta}^{(t)}_{\mathrm{new}}
+
(1-\rho)\, \boldsymbol{\theta}^{(t-1)},
\end{equation}
where $\rho = 0.9$ is the learning rate, $\boldsymbol{\theta}^{(t)}_{\mathrm{new}}$ denotes the MFVB update obtained from the current coordinate ascent step, and $\boldsymbol{\theta}^{(t-1)}$ is the estimate from the previous iteration. This under-relaxation scheme improves convergence behavior and reduces oscillations in high-dimensional settings.

\subsection{Model, Priors, and Hyperparameters}
For completeness, we summarize the model specifications, prior distributions, and associated hyperparameters used in the posterior computation. For each gene $i$ and FOV $k$, conditional on 
$\{u_{ijl}^k, \delta_{ij}^k, \omega_{ii}^k\}$,
the observation vector $Y_i(\mathbb{S}^k)$ follows
\begin{align*}
Y_{i}(\mathbb{S}^k) \mid \{u_{ijl}^k, \delta_{ij}^k, \omega_{ii}^k\}
\sim \mathcal{N}\!\left(
\sum_{j \neq i} Y_{j}(\mathbb{S}^k) \otimes \gamma_{ij}(\mathbb{S}^k)\, I(z_{ij}^k > 0),\;
(\omega_{ii}^k)^{-1} I_{N_k}
\right).
\end{align*}

\paragraph{Residual precision:}
$\omega_{ii}^k \sim \text{Gamma}(a_{\omega}, b_{\omega}).$

\paragraph{GP coefficients (KL-based):}
For $\tilde{v}_{ijl}^k = u_{ijl}^k \sqrt{\eta_l}, 
\tilde{v}_{ijl}^k \sim N(0,\; \sigma_{gp}^2 \eta_l).
$

\paragraph{Edge-selection latent variables (probit):}
\begin{align*}
z_{ij}^k &= \lambda_{ij}^k + \epsilon_{ij}^k, \qquad
\epsilon_{ij}^k \sim N(0,1), \\
\delta_{ij}^k &= I(z_{ij}^k > 0), \qquad
P(\delta_{ij}^k = 1) = \Phi(\lambda_{ij}^k).
\end{align*}

\paragraph{Spatial proximity prior on structured parameters:}
$\Lambda_{ij} \sim \mathcal{N}\!\left(m_{\Lambda i,\text{prior}},\; \sigma_{\Lambda i}^2 U\right).$

\paragraph{Hyperparameters:}
Throughout, the hyperparameters 
$a_{\omega},\; b_{\omega},\; \sigma_{gp}^2,\; 
m_{\Lambda i,\text{prior}},\; \sigma_{\Lambda i}^2,\; U
$
are treated as known and fixed. In simulation studies, we set
$a_w = 10, b_w=10, m_{\Lambda i,\text{prior}}, = 0, \sigma_{\Lambda i}^{-2}= 50.$ For GP parameters, we set $\sigma_{gp}^2 = 1, a_{gp} = 0.01, b
_{gp} = 0.5, $ approximation degree $=10,$ leading to $L =66$.

\paragraph{Notational details.}
For field of view (FOV) $k$ with $N_k$ spatial locations and a truncated basis expansion of size $L$, we define the following quantities.

\begin{enumerate}
    \item \textbf{Design matrix.}  
    For each pair $(i,j)$, let
    \[
    H_{ij}^k \in \mathbb{R}^{N_k \times L},
    \]
    where the $l^{\text{th}}$ column is given by
    \[
    \big(H_{ij}^k\big)_{\cdot l}
    = Y_j(\mathbb{S}^k) \otimes \psi_l(\mathbb{S}^k).
    \]

    \item \textbf{Scaled basis coefficients.}  
    Define the scaled coefficients
    \[
    \tilde{v}_{ijl}^k := u_{ijl}^k \sqrt{\eta_l},
    \qquad \tilde{v}_{ijl}^k \sim \mathcal{N}(0,\eta_l),
    \]
    for $l = 1,\dots,L$.

    \item \textbf{Coefficient vector.}  
    Collect the scaled coefficients into the vector
    \[
    \tilde{v}_{ij}^k
    = \big(\tilde{v}_{ij1}^k,\dots,\tilde{v}_{ijL}^k\big)^{\mathsf T}
    \in \mathbb{R}^L.
    \]

    \item \textbf{Basis expansion of spatial interaction.}  
    The spatially varying regression term admits the representation
    \[
    Y_j(\mathbb{S}^k) \otimes \gamma_{ij}(\mathbb{S}^k)
    = \sum_{l=1}^L
    \big(Y_j(\mathbb{S}^k) \otimes \psi_l(\mathbb{S}^k)\big)\,
    \tilde{v}_{ijl}^k
    = H_{ij}^k \tilde{v}_{ij}^k
    \in \mathbb{R}^{N_k}.
    \]

    \item \textbf{Partial residual.}  
    Define the partial residual excluding the $(i,j)$ interaction as
    \[
    R_{ij}^k
    = Y_i(\mathbb{S}^k)
    - \sum_{j' \neq i,j}
    \big\{
    Y_{j'}(\mathbb{S}^k) \otimes \gamma_{ij'}(\mathbb{S}^k)
    \big\}
    \, I\!\left(z_{ij'}^k > 0\right).
    \]

    \item \textbf{Quadratic form for the prior.}  
    The Gaussian prior penalty can be written as
    \[
    \sum_{l=1}^L \frac{(\tilde{v}_{ijl}^k)^2}{2\sigma_{ijk}^2 \eta_l}
    =
    \frac{1}{2\sigma_{ijk}^2}
    (\tilde{v}_{ij}^k)^{\mathsf T}
    \operatorname{diag}\!\left(
    \frac{1}{\eta_1},\dots,\frac{1}{\eta_L}
    \right)
    \tilde{v}_{ij}^k,
    \]
    where we denote
    \[
    \hat{\eta}^{-1}
    := \operatorname{diag}\!\left(
    \frac{1}{\eta_1},\dots,\frac{1}{\eta_L}
    \right).
    \]
\end{enumerate}

\subsection{Initialization of Spatial Coefficients} \label{suppsec:initialization}
We initialize the random variables $u_{ijl}^k$ in a field-of-view (FOV)--specific manner. For each FOV $k = 1,\ldots,K$, we extract the corresponding subsets of the response vector $\mathbf{Y}_i(\mathbb{S}^k)$, the predictor matrix $\mathbf{Y}_{-i}(\mathbb{S}^k)$, and the spatial basis matrix $B(\mathbb{S}^k) := \left[B_1(\mathbb{S}^k) \cdots B_L(\mathbb{S}^k)\right]$. A joint design matrix is then constructed by element-wise multiplying each predictor $\mathbf{Y}_{-i}(\mathbb{S}^k)$ with the spatial basis functions, yielding a matrix of dimension $n_k \times \{L(p-1)\}$. Using this joint design matrix, we fit a normal linear model via mean-field variational Bayes to obtain posterior mean estimates of the regression coefficients. These estimates are reshaped into an $L \times (p-1)$ matrix and used as the initial values for the coefficient matrix $u_{ij}^{k}$.

\subsection{Likelihood}\label{suppsec:msgr_likelihood}

$\begin{aligned}
L = &\prod_{i=1}^p \prod_{k=1}^k \left\{ \det([\omega_{ii}^k]^{-1} I_{N_k})^{-\frac{1}{2}} \exp\left(-\frac{\omega_{ii}^k}{2} \|Y_{i}(\mathbb{S}^k) - \sum_{j \neq i}^p Y_{j}(\mathbb{S}^k) \otimes \gamma_{ij}(\mathbb{S}^k) \cdot I(z_{ij}^k > 0) \|^2 \right) \right. \\
&\hspace{4em} \left.
 (\omega_{ii}^k)^{a_{\omega}-1} \exp(-b_{\omega} \cdot \omega_{ii}^k) \right\} \\
&\prod_{i=1}^p \prod_{j \neq i} ^p \prod_{k=1}^k \prod_{l=1}^L \left\{ (\sigma_{gp}^2 \eta^l)^{-\frac{1}{2}} \exp\left(-\frac{(\tilde{v}_{ijl}^k)^2}{2 \sigma_{gp}^2 \eta^l}\right) \right\}\\
&\prod_{i=1}^p \prod_{j \neq i} ^p \prod_{k=1}^k \left\{
 \exp\left(-\frac{1}{2} (z_{ij}^k - \lambda_{ij}^k)^2 \right)
\right\} \\
&\prod_{i=1}^p \prod_{j \neq i}^p \left\{ \exp\left(-\frac{1}{2} (\Lambda_{ij} - m_{\Lambda i, \text{prior}})^T (\sigma_{\Lambda i}^2 U)^{-1} (\Lambda_{ij} - m_{\Lambda i, \text{prior}}) \right) \right\}. \\
\end{aligned}$


\subsection{Augmented Mean Field Variational Bayes (MFVB)}\label{suppsec:msgr_MFVB}
\begin{enumerate}
    \item Update $\omega_{ii}^k$:
    \[
    \omega_{ii}^k \sim \text{Gamma}\left( \frac{N_k}{2} + a_{\omega}, \frac{1}{2} \mathbb{E} \left\| Y_{i}(\mathbb{S}^k) - \sum_{j \neq i} H_{ij}^k \cdot \tilde v_{ij}^k \cdot I(z_{ij}^k > 0) \right\|^2 + b_{\omega} \right).
    \]

    \item Update $\Lambda_{ij}$:
    \[
    \begin{aligned}
    \mathbb{E}_{-\Lambda_{ij}} l = &\mathbb{E}_{-\Lambda_{ij}} \left\{- \frac{1}{2} \left( \Lambda_{ij} - Z_{ij} \right)^T \left( \Lambda_{ij} - Z_{ij} \right) - \right. \\
    &\hspace{3em} \left. 
    \frac{1}{2 \sigma_{\Lambda i}^2} \left( \Lambda_{ij} - m_{\Lambda i, \text{prior}} \right)^T U^{-1} \left( \Lambda_{ij} - m_{\Lambda i, \text{prior}} \right)
    \right\}
    \end{aligned}
    \]
    \[
    \begin{aligned}
        \Lambda_{ij} \sim \mathcal{N} &\left( \left( I + \frac{1}{\sigma_{\Lambda i}^2}U^{-1} \right)^{-1} \left( \mathbb{E} Z_{ij} + \frac{1}{\sigma_{\Lambda i}^2}U^{-1} m_{\Lambda i, \text{prior}} \right),  \left( I + \frac{1}{\sigma_{\Lambda i}^2}U^{-1} \right)^{-1} \right).
    \end{aligned}
    \]

   \item Update $\tilde v_{ij}^k \mid z_{ij}^k$:
\[
\begin{aligned}
\log q(\tilde v_{ij}^k, z_{ij}^k)
&= \mathbb{E}_{-(\tilde v_{ij}^k, z_{ij}^k)}
\Bigg[
-\frac{\omega_{ii}^k}{2}
\Big\|
Y_i(\mathbb{S}^k)
- \sum_{j \neq i} H_{ij}^k \tilde v_{ij}^k I(z_{ij}^k>0)
\Big\|^2  \\
&\qquad
- \frac{1}{2} (z_{ij}^k - \lambda_{ij}^k)^2
- \frac{1}{2\sigma_{gp}^2}
(\tilde v_{ij}^k)^{\!T} \hat{\eta}^{-1} \tilde v_{ij}^k
\Bigg], \\
\\
\log q(\tilde v_{ij}^k \mid z_{ij}^k)
&= \mathbb{E}_{-\tilde v_{ij}^k \mid z_{ij}^k}
\Bigg[
-\frac{\omega_{ii}^k}{2}
\Big\|
Y_i(\mathbb{S}^k)
- \sum_{j' \neq i} H_{ij'}^k \tilde v_{ij'}^k I(z_{ij'}^k>0)
\Big\|^2  \\
&\qquad
- \frac{1}{2\sigma_{gp}^2}
(\tilde v_{ij}^k)^{\!T} \hat{\eta}^{-1} \tilde v_{ij}^k
\Bigg], \\
&= \mathbb{E}_{-\tilde v_{ij}^k \mid z_{ij}^k}
\Bigg[
-\frac{\omega_{ii}^k}{2}
\big\|
R_{ij}^k
- H_{ij}^k \tilde v_{ij}^k I(z_{ij}^k>0)
\big\|^2
- \frac{1}{2\sigma_{gp}^2}
(\tilde v_{ij}^k)^{\!T} \hat{\eta}^{-1} \tilde v_{ij}^k
\Bigg].
\end{aligned}
\]
Thus,
\[
\tilde v_{ij}^k \mid z_{ij}^k
\sim
\mathcal{MVN}\!\left(
\Hat{\mu}_{\tilde v}(z_{ij}^k),
\Hat{\Sigma}_{\tilde v}(z_{ij}^k)
\right).
\]
    with
\[
\Hat{\Sigma}_{\tilde v} (z_{ij}^k) = \left[ \mathbb{E} \omega_{ii}^k I(z_{ij}^k > 0) \|H_{ij}^k\|^2 + \mathbb{E} \frac{1}{\sigma_{gp}^2} \hat \eta^{-1} \right]^{-1},
\]
\[
\Hat{\mu}_{\tilde v} (z_{ij}^k) = \Hat{\Sigma}_{\tilde v} (z_{ij}^k) \cdot \mathbb{E} {\omega_{ii}^k} \left[ I(z_{ij}^k > 0) \right] (H_{ij}^k)^T \mathbb{E} R_{ij}^k.
\]

        \item Update $z_{ij}^k$:
\[
\begin{aligned}
   &q(\tilde v_{ij}^k, z_{ij}^k) \\
   =& \exp \left\{ \mathbb{E} \left[-\frac{\omega_{ii}^k}{2} \| R_{ij}^k - H_{ij}^k \tilde v_{ij}^k \cdot I(z_{ij}^k>0) \|^2 - \frac{1}{2\sigma_{gp}^2}  (\tilde v_{ij}^k)^T \hat \eta^{-1} \tilde v_{ij}^k - \frac{1}{2} (z_{ij}^k - \lambda_{ij}^k)^2 \right]\right\} \\
    \\
    &q(z_{ij}^k) = \int q(\tilde v_{ij}^k, z_{ij}^k) d \tilde v_{ij}^k \\
    =& \det(\Hat{\Sigma}_{\tilde v} (z_{ij}^k))^{1/2} \exp \left\{ \frac{1}{2} \Hat{\mu}_{\tilde v}^T (z_{ij}^k) \Hat{\Sigma}^{-1}_{\tilde v} (z_{ij}^k) \Hat{\mu}_{\tilde v} (z_{ij}^k) \right\} \exp \left\{ -\frac{1}{2} (z_{ij}^k)^2 + \mathbb{E} \lambda_{ij}^k z_{ij}^k \right\} \\
    &\sim p_{ij}^k \cdot TN(\mathbb{E} \lambda_{ij}^k,1,0,+\infty) + (1-p_{ij}^k) \cdot TN(\mathbb{E} \lambda_{ij}^k,1,-\infty,0) \\
    \\
    &p_{ij}^k = q(z_{ij}^k > 0) = \int_{0}^{+\infty} q(z_{ij}^k) dz_{ij}^k\\
    =& \det(\Hat{\Sigma}_{\tilde v} (z_{ij}^k>0))^{1/2} \cdot \exp \left\{ \frac{1}{2} \Hat{\mu}_{\tilde v}^T (z_{ij}^k>0) \Hat{\Sigma}^{-1}_{\tilde v} (z_{ij}^k>0) \Hat{\mu}_{\tilde v} (z_{ij}^k>0) \right\}  \\
    & \quad \Phi(\mathbb{E} \lambda_{ij}^k) \cdot \text{Const} \\
    \\
    &1-p_{ij}^k = q(z_{ij}^k < 0) = \int_{-\infty}^{0} q(z_{ij}^k) dz_{ij}^k\\
    =& \det(\Hat{\Sigma}_{\tilde v} (z_{ij}^k<0))^{1/2} \cdot [1-\Phi(\mathbb{E} \lambda_{ij}^k)] \cdot \text{Const} \\
    \\
    & \text{logit}(p_{ij}^k) = \frac{1}{2}\left[
    \log \det(\Hat{\Sigma}_{\tilde v} (z_{ij}^k>0)) -
    \log \det(\Hat{\Sigma}_{\tilde v} (z_{ij}^k<0))
    \right] + \\
    &\quad \frac{1}{2} \Hat{\mu}_{\tilde v}^T (z_{ij}^k>0) \Hat{\Sigma}^{-1}_{\tilde v} (z_{ij}^k>0) \Hat{\mu}_{\tilde v} (z_{ij}^k>0) + \text{logit}(\Phi(\mathbb{E} \lambda_{ij}^k)).
\end{aligned}
\]
        
    \end{enumerate}

\
\textbf{\hspace{-1.75em} Numerical note:}
When $\mathbb{E}[\lambda_{ij}^k]$ is positively or negatively large (e.g., $|\mathbb{E}[\lambda_{ij}^k]| > 10$), we have $\Phi(\mathbb{E}[\lambda_{ij}^k]) \to 1$ or $0$, and hence $\text{logit}(\Phi(\mathbb{E}[\lambda_{ij}^k]))$ is not well-defined numerically. Using Mills' ratio $\lim_{x \to \infty} \frac{1 - \Phi(x)}{\frac{1}{\sqrt{2\pi}} e^{-x^2/2}} = \frac{1}{x}$, we obtain the following approximations:
\begin{enumerate}
    \item When $x$ is positively large, $1 - \Phi(x) \approx \frac{1}{\sqrt{2\pi}} \cdot \frac{e^{-x^2/2}}{x}$. Thus
    \[
    \log \left( \frac{\Phi(x)}{1 - \Phi(x)} \right) \approx \frac{1}{2} \log 2\pi + \log x + \frac{x^2}{2}.
    \]
    \item When $x$ is negatively large, $\Phi(x) \approx -\frac{1}{\sqrt{2\pi}} \cdot \frac{e^{-x^2/2}}{x}$, and
    \[
    \log \left( \frac{\Phi(x)}{1 - \Phi(x)} \right) \approx -\frac{1}{2} \log 2\pi - \log (-x) - \frac{x^2}{2}.
    \]
\end{enumerate}

\paragraph{Expression of expected values:}
\begin{enumerate}
    \item $\mathbb{E} \| Y_{i}(\mathbb{S}^k) - \sum_{j \neq i}^p H_{ij}^k \tilde{v}_{ij}^k \cdot I(z_{ij}^k > 0) \|^2$  

    Denote $G_{ijk} = \mathbb{E}(H_{ij}^k \tilde{v}_{ij}^k \cdot I(z_{ij}^k > 0)) = p_{ij}^k H_{ij}^k \Hat{\mu}_{\tilde{v}} (z_{ij}^k>0)$ and $\Tilde{G}_{ik} = \sum_{j \neq i}^p G_{ijk}$. Then
    \[
    \begin{aligned}
        &\mathbb{E} \| Y_{i}(\mathbb{S}^k) - \sum_{j \neq i}^p H_{ij}^k \tilde{v}_{ij}^k \cdot I(z_{ij}^k > 0) \|^2 \\
        =& \| Y_{i}(\mathbb{S}^k) \|^2 + \sum_{j \neq i}^p \mathbb{E} \| H_{ij}^k \tilde{v}_{ij}^k \cdot I(z_{ij}^k > 0) \|^2 - 2Y^T_{ik}(S)\mathbb{E}\left[ \sum_{j \neq i}^p H_{ij}^k \tilde{v}_{ij}^k \cdot I(z_{ij}^k > 0)\right] +\\
        & \quad \sum_{j \neq i}^p \sum_{j' \neq i,j}^p \mathbb{E}\left[I(z_{ij}^k > 0) (\tilde{v}_{ij}^k)^T (H_{ij}^k)^T H_{ij'}k \tilde{v}_{ij'}^k I(z_{ij'k} > 0) \right].
    \end{aligned}
    \]

    \begin{enumerate}
        \item $\mathbb{E} \| H_{ij}^k \tilde{v}_{ij}^k \cdot I(z_{ij}^k > 0) \|^2$:
        \begin{align*}
           &\mathbb{E} \| H_{ij}^k \tilde{v}_{ij}^k \cdot I(z_{ij}^k > 0) \|^2 \\
           =& p_{ij}^k \left[\Hat{\mu}_{\tilde{v}}^T (z_{ij}^k>0) ([H_{ij}^k]^T H_{ij}^k) \Hat{\mu}_{\tilde{v}} (z_{ij}^k>0) + \tr \left([H_{ij}^k]^T H_{ij}^k \Hat{\Sigma}_{\tilde{v}}(z_{ij}^k>0) \right)\right] \\
           =& \frac{1}{p_{ij}^k} G_{ijk}^T G_{ijk} + p_{ij}^k \cdot \tr \left([H_{ij}^k]^T H_{ij}^k \Hat{\Sigma}_{\tilde{v}}(z_{ij}^k>0) \right).
        \end{align*}
        \item $Y_{ik}^T(S) \cdot \mathbb{E} \left[
        \sum_{j \neq i} H_{ij}^k \tilde{v}_{ij}^k \cdot I(z_{ij}^k >0)
        \right] = Y_{ik}^T(S) \cdot \Tilde{G}_{ik}$.
        \item $\sum_{j \neq i}^p \sum_{j' \neq i,j}^p \mathbb{E}\left[I(z_{ij}^k > 0) (u_{ij}^k)^T (H_{ij}^k)^T H_{ij'k} u_{ij'k} I(z_{ij'k} > 0)\right]$:
        \begin{align*}
            &\sum_{j \neq i}^p \sum_{j' \neq i,j}^p \mathbb{E}\left[I(z_{ij}^k > 0) (u_{ij}^k)^T (H_{ij}^k)^T H_{ij'k} u_{ij'k} I(z_{ij'k} > 0)\right] \\
        =& \sum_{j \neq i}^p \sum_{j' \neq i,j}^p\left[G_{ijk}^T G_{ij'k}\right] = \sum_{j \neq i}^p \left[G_{ijk}^T (\Tilde{G}_{ik} - G_{ijk})\right] = \Tilde{G}_{ik}^T\Tilde{G}_{ik} - \sum_{j \neq i}^p G_{ijk}^T G_{ijk}.
        \end{align*}
    \end{enumerate}

    \item $\mathbb{E} [(\tilde{v}_{ij}^k)^T \tilde{v}_{ij}^k]$:
    \begin{align*}
        &\mathbb{E}[(\tilde{v}_{ij}^k)^T \tilde{v}_{ij}^k] \\
        =& \mathbb{E}_{z_{ij}^k}\left\{\mathbb{E} [(\tilde{v}_{ij}^k)^T \tilde{v}_{ij}^k|z_{ij}^k)]\right\} \\
        =&\mathbb{E}_{z_{ij}^k}\left[ \mathbb{E}^T(\tilde{v}_{ij}^k \mid z_{ij}^k) \mathbb{E}(\tilde{v}_{ij}^k \mid z_{ij}^k) + \text{tr}(\text{Var}(\tilde{v}_{ij}^k \mid z_{ij}^k)) \right] \\
        =&p_{ij}^k \cdot \left[ \Hat{\mu}_{\tilde{v}} (z_{ij}^k>0)^T \Hat{\mu}_{\tilde{v}} (z_{ij}^k>0) + \text{tr}(\Hat{\Sigma}_{\tilde{v}}(z_{ij}^k>0)) \right] + (1-p_{ij}^k) \text{tr}(\Hat{\Sigma}_{\tilde{v}}(z_{ij}^k<0)).
    \end{align*}
    
    \item $\mathbb{E} \lambda_{ij}^k = \mu_{\lambda_{ij}^k}$.

    \item $\mathbb{E} z_{ij}^k = \mathbb{E} \lambda_{ij}^k + p_{ij}^k \frac{\phi(\mathbb{E} \lambda_{ij}^k)}{\Phi(\mathbb{E} \lambda_{ij}^k)} - (1-p_{ij}^k)\frac{\phi(\mathbb{E} \lambda_{ij}^k)}{1-\Phi(\mathbb{E} \lambda_{ij}^k)}$.

    Approximation:

    When $\mathbb{E} \lambda_{ij}^k \to +\infty$, we have $\frac{\phi(\mathbb{E} \lambda_{ij}^k)}{\Phi(\mathbb{E} \lambda_{ij}^k)} \approx 0$ and $\frac{\phi(\mathbb{E} \lambda_{ij}^k)}{1-\Phi(\mathbb{E} \lambda_{ij}^k)} \approx \mathbb{E} \lambda_{ij}^k$.

    When $\mathbb{E} \lambda_{ij}^k \to -\infty$, we have $\frac{\phi(\mathbb{E} \lambda_{ij}^k)}{\Phi(\mathbb{E} \lambda_{ij}^k)} \approx -\mathbb{E} \lambda_{ij}^k$ and $\frac{\phi(\mathbb{E} \lambda_{ij}^k)}{1-\Phi(\mathbb{E} \lambda_{ij}^k)} \approx 0$.

    \item $\mathbb{E}[\omega_{ii}^k] = \frac{n_k/2 + a_{\omega}}{\frac{1}{2} \mathbb{E}\|Y_{i}(\mathbb{S}^k) - \sum_{j \neq i}^p H_{ij}^k \cdot \tilde{v}_{ij}^k \cdot I(z_{ij}^k > 0)\|^2 + b_{\omega}}$.

    \item $\mathbb{E}[R_{ij}^k] = Y_{i}(\mathbb{S}^k) - \sum_{j' \neq i,j} G_{ij'k}$.
\end{enumerate}

\newpage
\subsection{Algorithm} \label{suppsec:algorithm}
\begin{algorithm}[H]
\caption{Variational inference for \texttt{mSGR}}
\label{alg:msgr_vb_clean}
\small
\begin{algorithmic}
\State \textbf{Input:} Data $\{Y_i(\mathbb{S}^k)\}$, $\{H_{ij}^k\}$, hyperparameters
$a_\omega,b_\omega,\sigma_{\Lambda i}^2,U,m_{\Lambda i,\mathrm{prior}},\sigma_{\mathrm{gp}}^2,\hat\eta$,
damping \texttt{step} $\in(0,1]$, tolerance \texttt{tol}, max iter $T_{\max}$
\
\State \textbf{Parameters:} $\omega_{ii}^k,\Lambda_{ij}, \tilde{v}_{ij}^k, p_{ij}^k, z_{ij}^k$
\
\State Initialize all variational parameters $\theta^{(0)}$.
\While $\|\theta^{(t)}-\theta^{(t-1)}\|_{\infty}> \texttt{tol } \& \ t\leq T_{\max}$ do
\For{$k=1,\dots,K$} 
  \For{$i=1,\dots,p$} 
    \State \textbf{1. Update} $\mathbb{E}_q(\omega_{ii}^k)$ with
    \Statex \hspace{\algorithmicindent} \hspace{\algorithmicindent} \hspace{\algorithmicindent}
    $
    \omega_{ii}^k \sim 
    \text{Gamma}\!\left(
        \frac{N_k}{2} + a_{\omega},\,
        \frac{1}{2}\,
        \mathbb{E}\left\|
        Y_{i}(\mathbb{S}^k)
        - \sum_{j \neq i}
        H_{ij}^k \cdot \tilde v_{ij}^k \cdot I(z_{ij}^k > 0)
        \right\|^2
        + b_{\omega}
    \right)
    $
    
    \For{$j\neq i$}
    \State \textbf{2. Update} $\mathbb{E}_q(\Lambda_{ij})$ with
\Statex \hspace{\algorithmicindent} \hspace{\algorithmicindent} \hspace{\algorithmicindent}
    $
    \Lambda_{ij} \sim \mathcal{N} \left[ \left( I + \frac{1}{\sigma_{\Lambda i}^2}U^{-1} \right)^{-1} \left( \mathbb{E} Z_{ij} + \frac{1}{\sigma_{\Lambda i}^2}U^{-1} m_{\Lambda i, \text{prior}} \right),  \left( I + \frac{1}{\sigma_{\Lambda i}^2}U^{-1} \right)^{-1} \right]
    $

    \State \textbf{3. Update} $\mathbb{E}_q(\tilde v_{ij}^k \mid z_{ij}^k)$ with
    $\tilde v_{ij}^k \mid z_{ij}^k \sim \mathcal{MVN} \left( \Hat{\mu}_{\tilde v} (z_{ij}^k), \Hat{\Sigma}_{\tilde v} (z_{ij}^k)\right)$
    \Statex \hspace{\algorithmicindent} \hspace{\algorithmicindent} \hspace{\algorithmicindent}\hspace{\algorithmicindent} 
    $
\Hat{\Sigma}_{\tilde v} (z_{ij}^k) = \left[ \mathbb{E} \omega_{ii}^k I(z_{ij}^k > 0) \|H_{ij}^k\|^2 + \mathbb{E} \frac{1}{\sigma_{gp}^2} \hat \eta^{-1} \right]^{-1}$
\Statex \hspace{\algorithmicindent} \hspace{\algorithmicindent} \hspace{\algorithmicindent}\hspace{\algorithmicindent} 
    $
\Hat{\mu}_{\tilde v} (z_{ij}^k) = \Hat{\Sigma}_{\tilde v} (z_{ij}^k) \cdot \mathbb{E} {\omega_{ii}^k} \left[ I(z_{ij}^k > 0) \right] (H_{ij}^k)^T \mathbb{E} R_{ij}^k$

      \State \textbf{4. Update} $\mathbb{E}z_{ij}^k$ with
      \Statex \hspace{\algorithmicindent} \hspace{\algorithmicindent} \hspace{\algorithmicindent}\hspace{\algorithmicindent} 
      $z_{ij} \sim p_{ij}^k \cdot TN(\mathbb{E} \lambda_{ij}^k,1,0,+\infty) + (1-p_{ij}^k) \cdot TN(\mathbb{E} \lambda_{ij}^k,1,-\infty,0)$

      \State \textbf{5. Update} $p_{ij}^k$ with
      \Statex \hspace{\algorithmicindent} \hspace{\algorithmicindent} \hspace{\algorithmicindent}\hspace{\algorithmicindent} 
      $
      \text{logit}(p_{ij}^k) = \frac{1}{2}\left[
    \log \det(\Hat{\Sigma}_{\tilde v} (z_{ij}^k>0)) -
    \log \det(\Hat{\Sigma}_{\tilde v} (z_{ij}^k<0))
    \right] + $
    \Statex \hspace{12em} 
    $\frac{1}{2} \Hat{\mu}_{\tilde v}^T (z_{ij}^k>0) \Hat{\Sigma}^{-1}_{\tilde v} (z_{ij}^k>0) \Hat{\mu}_{\tilde v} (z_{ij}^k>0) + \text{logit}(\Phi(\mathbb{E} \lambda_{ij}^k))
    $
    \Statex \hspace{\algorithmicindent} \hspace{\algorithmicindent} \hspace{\algorithmicindent}\hspace{\algorithmicindent} Approximation is used when $\mathbb{E} \lambda_{ij}^k$ has large positive or negative value

    \EndFor
  \EndFor
\EndFor
\Statex \vspace{0.2em}
\Statex \textbf{Damping (relaxation) update}
\State Collect current parameters into $\theta^{(t)}$.
\State Apply
$\theta^{(t)} \leftarrow \texttt{step}\cdot\theta^{(t)} + (1-\texttt{step})\cdot\theta^{(t-1)}.$

\Statex \textbf{Convergence check}
\If{$\|\theta^{(t)}-\theta^{(t-1)}\|_{\infty}<\texttt{tol}$}
  \State \textbf{stop}
\EndIf
\EndWhile
\State \Return $\theta$.
\end{algorithmic}
\end{algorithm}

\newpage

\subsection{Postprocessing Steps}\label{suppsec:msgr_post_steps}

\paragraph{Selection threshold determination}: As discussed in Section \ref{sec:mfvb}, we use a Bayesian local false discovery rate (FDR) control procedure that aims to control the average Bayesian FDR at a pre-specified level $\alpha$ \citep{baladandayuthapani2010bayesian} to decide selection threshold $\kappa_\alpha$. Define the Bayesian q-value (or local FDR estimate) for each edge as $\hat{q}_{ij}^k = \hat{q}_{ji}^k = 1 - \hat{p}_{ij}^k$. The threshold $\kappa_\alpha$ is then obtained via the following procedure: 1) Sort the set $\{\hat{q}_{ij}^k\}_{1 \leq i < j \leq p,\, 1 \leq k \leq K}$ in ascending order and denote the ordered sequence by $\{\hat{q}^{(t)}\}_{1 \leq t \leq p(p-1)K/2}$; 2) Compute the cumulative average $\bar{q}^{(t)} = \frac{1}{t} \sum_{s=1}^t \hat{q}^{(s)}$; 3) Find the largest index $t^*$ such that $\bar{q}^{(t^*)} \leq \alpha$; 4) Set the threshold $\kappa_\alpha = 1 - \hat{q}^{(t^*)}$.

\paragraph{Positive definiteness property: }

From the GGM property, we have $\omega_{ij}(s^k)= - \omega_{ii}^k \cdot \gamma_{ij}(s^k)$. Based on Equations~\ref{eq:gamma_decompose} and~\ref{eq:gp_main}, for a selected edge we have
\[
\sum_{j\neq i}^p |\omega_{ij}(s^k)|
\approx
\omega_{ii}^k \sum_{j\neq i}^p \left|\sum_{l=1}^L u_{ijl}^k B_l(s^k)\right|.
\]
To ensure diagonal dominance, it suffices to require
\[
\sum_{j\neq i}^p |\omega_{ij}(s^k)| \le |\omega_{ii}^k|,
\qquad \forall s^k \in \mathbb{S}^k,
\]
which implies
\[
\sum_{j\neq i}^p \left|\sum_{l=1}^L u_{ijl}^k B_l(s^k)\right| \le 1,
\qquad \forall s^k \in \mathbb{S}^k.
\]
Define $B(\mathbb{S}^k) := \left[ B_1(\mathbb{S}^k) \cdots B_L(\mathbb{S}^k) \right] \in \mathbb{R}^{N_k \times L}$ and let $a_k = \|B(\mathbb{S}^k)\|_\infty$, which is an observed quantity depending on the spatial domain and GP parameters. Then,
\[
\sum_{j\neq i}^p \left|\sum_{l=1}^L u_{ijl}^k B_l(s^k)\right|
\le
a_k \sum_{j\neq i}^p \left|\sum_{l=1}^L u_{ijl}^k\right|
\le
a_k \sum_{j\neq i}^p \sum_{l=1}^L |u_{ijl}^k|.
\]
Therefore, diagonal dominance is guaranteed if
\[
a_k \sum_{j\neq i}^p \sum_{l=1}^L |u_{ijl}^k| \le 1.
\]
To enforce this constraint, we rescale $u_{ijl}^k$ by the factor
$a_k \sum_{j\neq i}^p \sum_{l=1}^L |u_{ijl}^k|$.

\section{Simulation Studies} \label{suppsec:simulation}
\subsection{Metrics for comparison} We use the following measures to compare with other methods (I) true positive rate (TPR); (II) false positive rate (FPR); (III) false discovery rate (FDR); (IV) Matthews correlation coefficient (MCC). MCC \citep{matthews1975comparison} measures the quality of binary classification, ranging from +1 (perfect classification) to -1 (total mismatch).

\subsection{Comparative Methods} 

We compare \texttt{mSGR} with the following methods (I) Spatial Graphical Regression (SGR, \cite{acharyya2025spatially}); (II) GraphR \citep{chen2025probabilistic}; (III)
Joint Graphical Lasso (JGL, \cite{danaher2014joint}); (IV) jointGHS \citep{lingjaerde2024scalable} and (V) graphical lasso (Glasso, \cite{friedman2008sparse}) method applied independently to each FOV.

\begin{itemize}
    \item \textbf{SGR}: The original \texttt{SGR} method is implemented using a Markov chain Monte Carlo (MCMC) algorithm. To ensure a fair computational comparison with variational methods, we derive a MFVB approximation for \texttt{SGR} and implement it using the \texttt{mSGR} package. Hyperparameters are set to $a_{\omega} = b_{\omega} = 10$, and the prior edge inclusion probability $\pi$ is fixed to match the true sparsity level in each simulation setting. For GP parameters, we set $\sigma_{gp}^2 = 1, a_{gp} = 0.01, b_{gp} = 0.5, $ approximation degree $=10$. The MFVB algorithm is run until convergence using the same learning rate and stopping criteria as \texttt{mSGR}.

    \item \textbf{GraphR}: The algorithm is implemented using the \texttt{GraphR} package with hyperparameters set to $a_\pi = 1$, $b_\pi = 4$, and $a_\tau = b_\tau = 0.005$, with a maximum of 2000 iterations and a convergence tolerance of $10^{-3}$. We apply \texttt{GraphR} separately to each FOV, using scaled $(x,y)$ coordinates as continuous external covariates. Since \texttt{GraphR} outputs cell-level edge inclusion probabilities (rather than FOV-level summaries), we construct an FOV-level edge indicator by declaring an edge present in a given FOV if it is selected in at least one cell within that FOV.
    
   \item \textbf{GGL}: The tuning parameters $\lambda_1$ and $\lambda_2$ are selected from a $20 \times 20$ grid, evenly spaced between 0.05 and 0.75 for both $\lambda_1$ and $\lambda_2$, using the approximated Akaike Information Criterion (AIC). Each FOV is treated as a distinct group, and spatial coordinate information is not incorporated in this method.

   \item \textbf{jointGHS}: The algorithm is implemented using the \texttt{jointGHS} package with \texttt{epsilon} $=0.001$ and \texttt{AIC\_eps} $=0.001$. Similar to GGL, each FOV is regarded as a separate group, and no spatial coordinate information is used in model fitting.
   
   \item \textbf{GLASSO}: The tuning parameter $\lambda$ is selected using the stability-based approach \cite{liu2010stability}. The graphical lasso model is fit separately for each FOV, and no spatial coordinate information is incorporated.
   
\end{itemize}

\subsection{Additional Simulation Results of Scenario I}\label{suppsec:additional_sim1}
We compare \texttt{mSGR} with competing methods under alternative sparsity levels (10\%). All methods are implemented using the same parameter settings as in Section \ref{sec:simStudy} to ensure a fair comparison. \texttt{mSGR} consistently outperforms or matches competing approaches in terms of graph recovery, demonstrating its robustness to increased network density.

\begin{figure}[ht!]
  \centering
  \includegraphics[width=\textwidth]{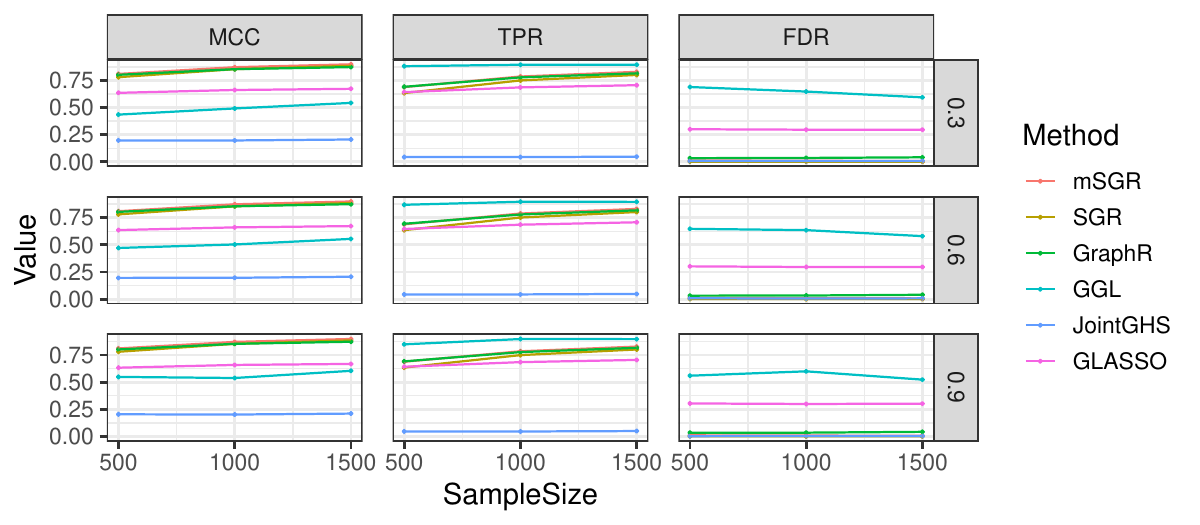}
  \caption{\small \textbf{Simulation performance of \texttt{mSGR}.} Simulation results comparing different methods under varying correlation strengths between FOVs ($\rho$ $=$ $0.3$, $0.6$, $0.9$) and sample sizes ($n=500,1000,1500$ per FOV) with sparsity fixed to be 10\%. The plot represents mean values of the  evaluation metrics (MCC, TPR, FDR, FPR) across replications.}
  \label{fig_supp:fig1}
\end{figure}

\begin{figure}[ht!]
  \centering
  \includegraphics[scale=0.5]{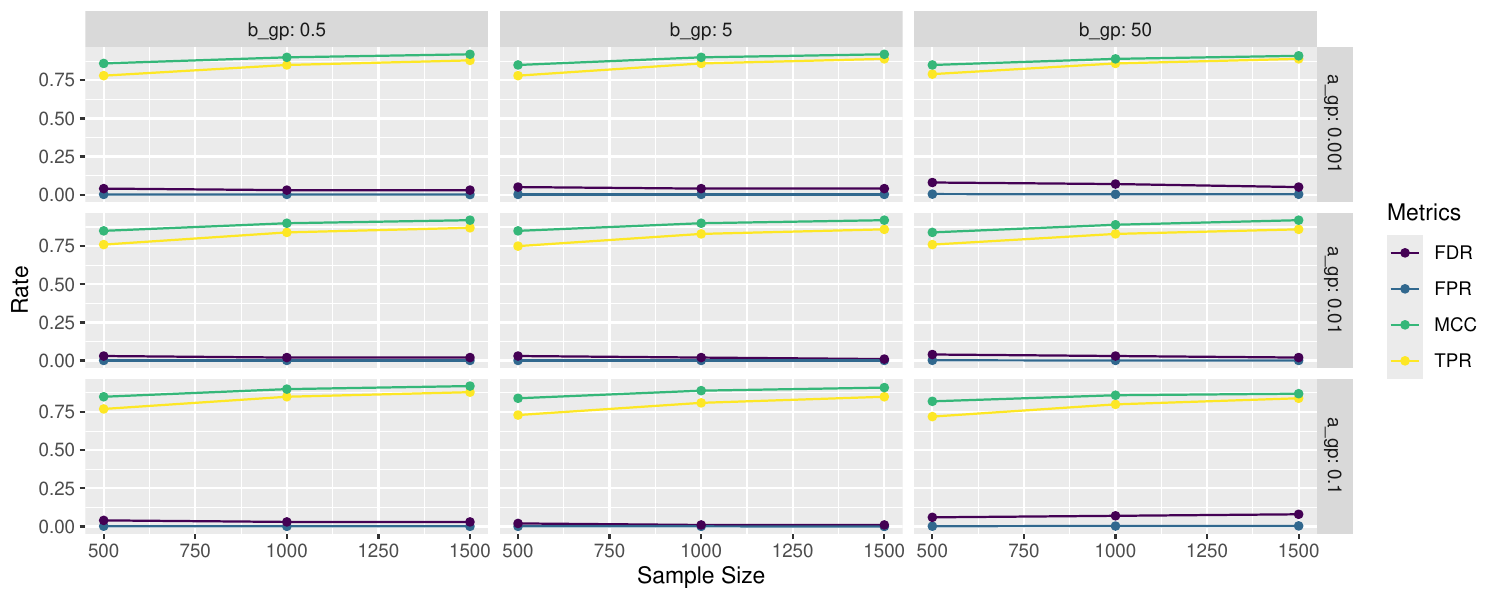}
  \caption{Simulation results comparing different hyperparameter settings, with $a_{gp} \in {0.001, 0.01, 0.1}$ and $b_{gp} \in {0.5, 5, 50}$. The correlation strength between FOVs is fixed at 0.3, sparsity is set to 5\%, and the per-FOV sample size varies across $n \in {500, 1000, 1500}$.
}  \label{fig_supp:fig2}
\end{figure}

\begin{figure}[ht!]
  \centering
  \includegraphics[scale=0.5]{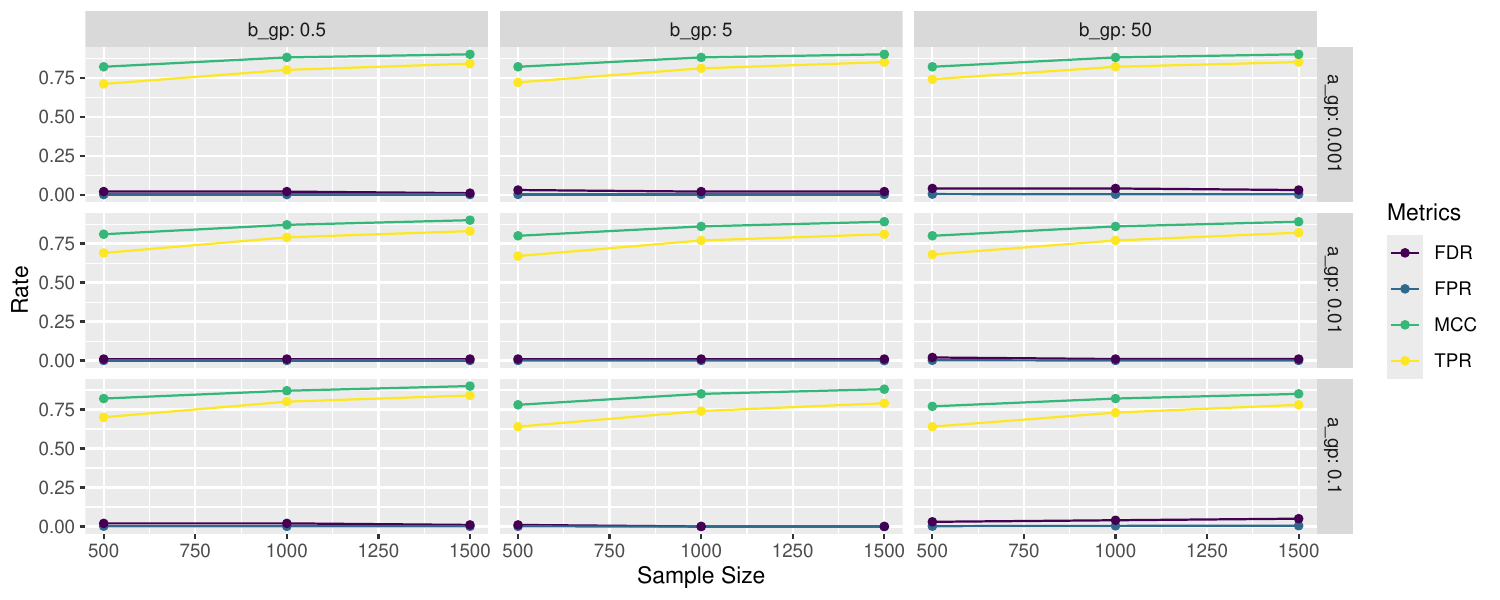}
  \caption{Simulation results comparing different hyperparameter settings, with $a_{gp} \in {0.001, 0.01, 0.1}$ and $b_{gp} \in {0.5, 5, 50}$. The correlation strength between FOVs is fixed at 0.3, sparsity is set to 10\%, and the per-FOV sample size varies across $n \in {500, 1000, 1500}$.
}\label{fig_supp:fig3}
\end{figure}

\begin{figure}[ht!]
  \centering
  \includegraphics[scale=0.5]{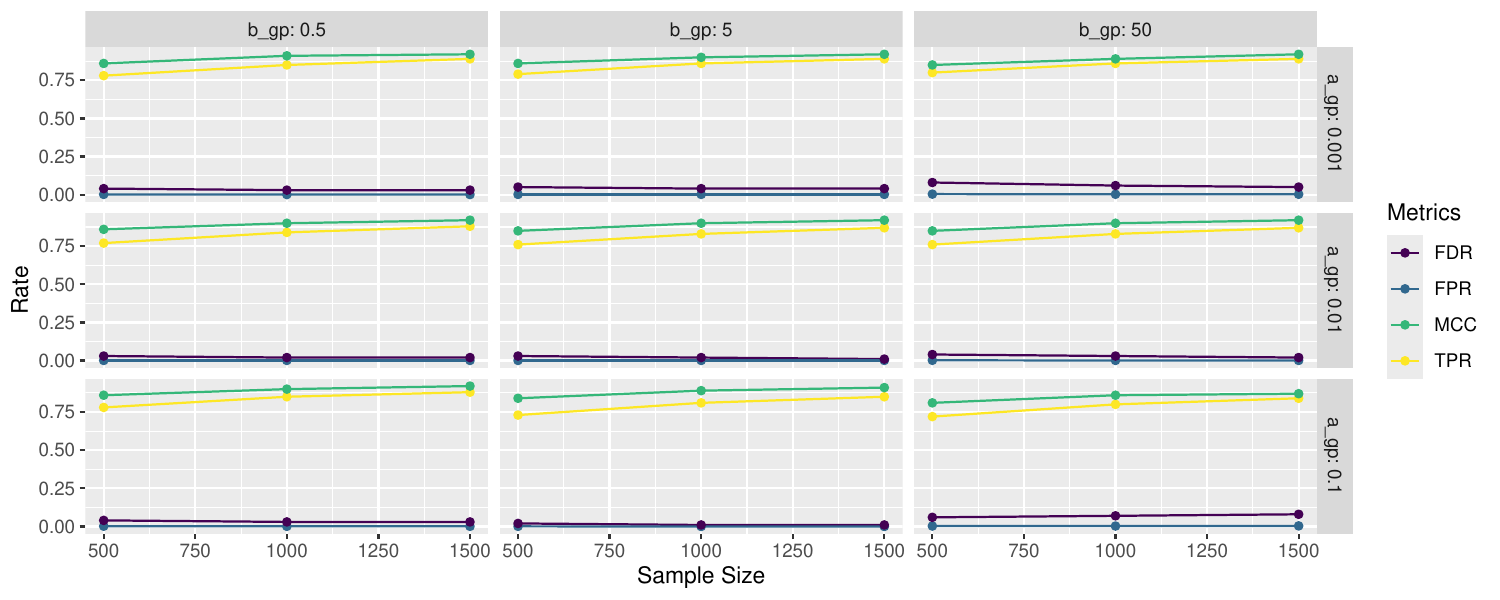}
  \caption{Simulation results comparing different hyperparameter settings, with $a_{gp} \in {0.001, 0.01, 0.1}$ and $b_{gp} \in {0.5, 5, 50}$. The correlation strength between FOVs is fixed at 0.6, sparsity is set to 5\%, and the per-FOV sample size varies across $n \in {500, 1000, 1500}$.
}\label{fig_supp:fig4}
\end{figure}

\begin{figure}[ht!]
  \centering
  \includegraphics[scale=0.5]{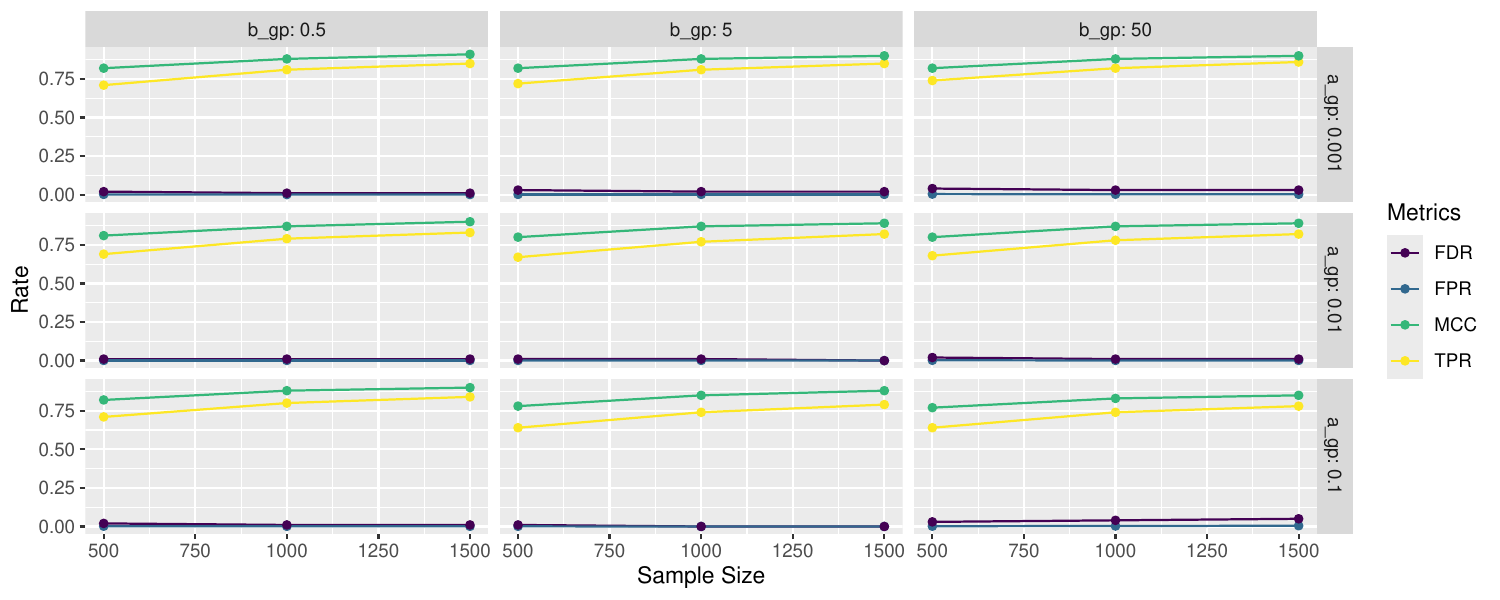}
  \caption{Simulation results comparing different hyperparameter settings, with $a_{gp} \in {0.001, 0.01, 0.1}$ and $b_{gp} \in {0.5, 5, 50}$. The correlation strength between FOVs is fixed at 0.6, sparsity is set to 10\%, and the per-FOV sample size varies across $n \in {500, 1000, 1500}$.
}\label{fig_supp:fig5}
\end{figure}

\begin{figure}[ht!]
  \centering
  \includegraphics[scale=0.5]{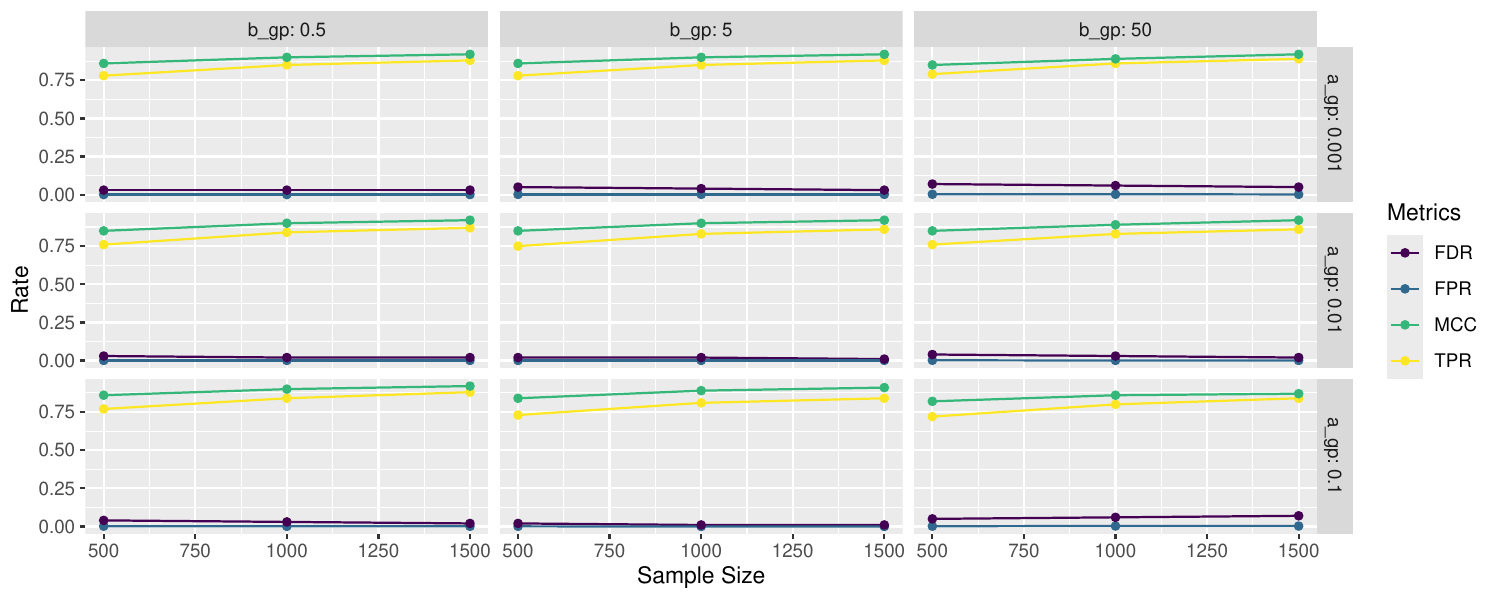}
  \caption{Simulation results comparing different hyperparameter settings, with $a_{gp} \in {0.001, 0.01, 0.1}$ and $b_{gp} \in {0.5, 5, 50}$. The correlation strength between FOVs is fixed at 0.9, sparsity is set to 5\%, and the per-FOV sample size varies across $n \in {500, 1000, 1500}$.
}\label{fig_supp:fig6}
\end{figure}

\begin{figure}[ht!]
  \centering
  \includegraphics[scale=0.5]{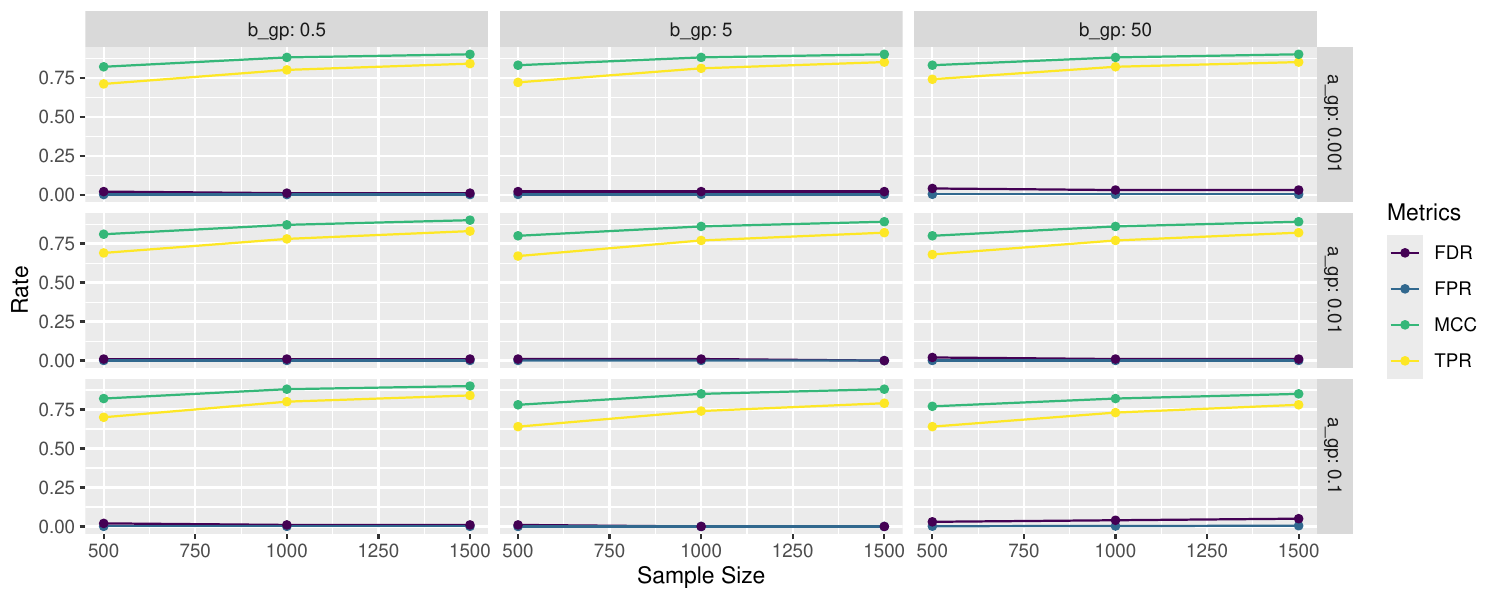}
  \caption{Simulation results comparing different hyperparameter settings, with $a_{gp} \in {0.001, 0.01, 0.1}$ and $b_{gp} \in {0.5, 5, 50}$. The correlation strength between FOVs is fixed at 0.9, sparsity is set to 10\%, and the per-FOV sample size varies across $n \in {500, 1000, 1500}$.
}\label{fig_supp:fig7}
\end{figure}

We evaluate the robustness of \texttt{mSGR} to the choice of Gaussian process hyperparameters by varying $a_{gp}$ and $b_{gp}$ over a wide range of values under different sparsity setting (5\% and 10\%) as shown in Figures \ref{fig_supp:fig1} $-$ \ref{fig_supp:fig7}. Across all configurations, \texttt{mSGR} exhibits stable performance in terms of edge recovery, with minimal sensitivity to the specific hyperparameter settings. These results suggest that the proposed method is robust to moderate misspecification of the GP prior.

\subsection{Computation Time}\label{suppsec:computation_time}
\begin{figure}[ht!]
  \centering
  \includegraphics[scale=0.8]{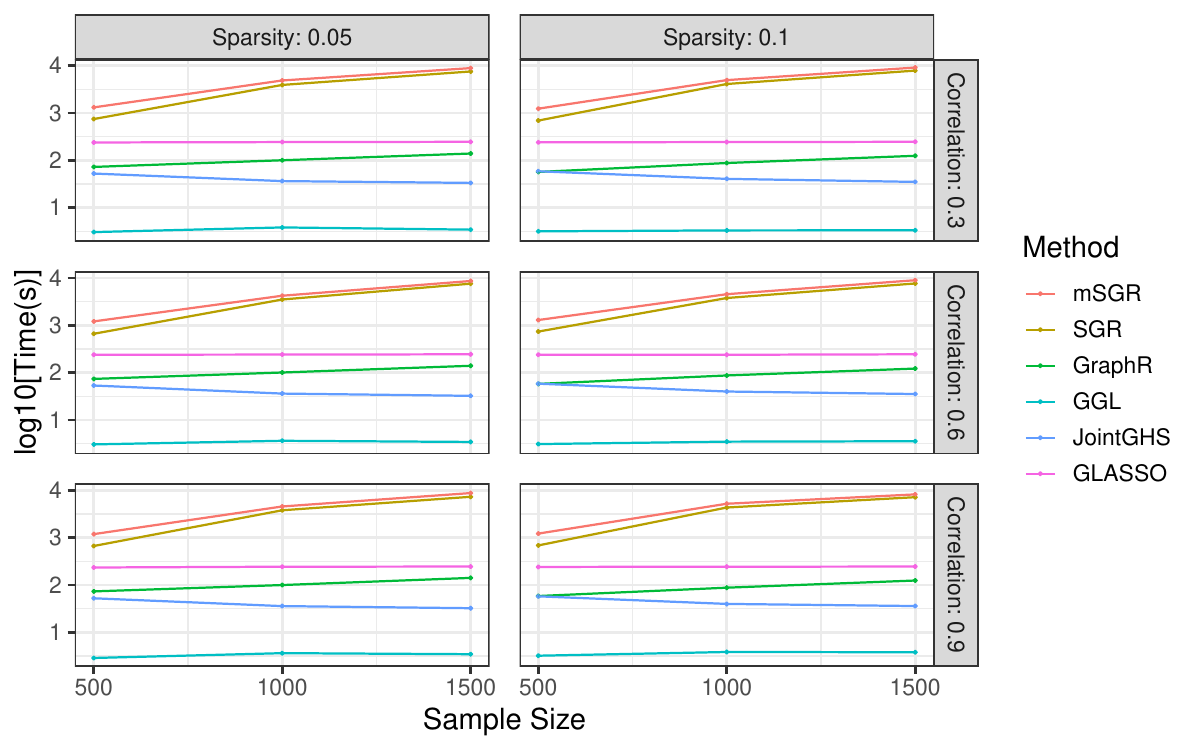}
  \caption{\textbf{Computation time comparison across methods for Scenario I.} Average computation time (log10 scale) as a function of sample size per FOV for \texttt{mSGR}, SGR, GraphR, GGL, JointGHS, and GLASSO. Results are shown for sparsity levels 0.05 and 0.1 (columns) and inter-FOV correlation strengths $\rho = 0.3, 0.6, 0.9$ (rows). Computation time is averaged across replicates and reported in second before log transformation.}
  \label{suppfig:computation_time_sim1}
\end{figure}

Figure \ref{suppfig:computation_time_sim1} summarizes the average computation time of all competing methods in Scenario I. Computation time increases with sample size for all methods, while remaining relatively insensitive to sparsity level and inter-FOV correlation strength. For \texttt{mSGR}, the average runtime per replicate increases from approximately 0.34 hours at $n=500$ to 1.26 hours at $n=1000$ and 2.43 hours at $n=1500$ per FOV, with similar trends observed across correlation and sparsity settings. Table~\ref{tab:supp_runtime_min} reports the average runtime per simulation replicate for each method in Scenario II, measured in minutes under the same computational environment using a single CPU core on a Linux-based HPC cluster.

\begin{table}[htbp]
\centering
\begin{tabular}{lc}
\toprule
Method & Time (min) \\
\midrule
\textbf{mSGR} & 80.17 \\
SGR & 57.19 \\
GraphR & 0.95 \\
GGL & 0.05 \\
JointGHS & 0.47 \\
GLASSO & 3.05 \\
\bottomrule
\end{tabular}
\caption{Average computation time per simulation replicate (in minutes) in Scenario II.}
\label{tab:supp_runtime_min}
\end{table}

\section{Spatial network 
characterization of EMT gradient in sarcomatoid renal cancer} \label{suppsec:application}
\subsection{Expression normalization} 

To prepare the gene expression data for spatial analysis, we exclude cells with total counts less than 10, resulting in 28,257 remaining cells. The raw count matrix is then normalized using the SCTransform procedure implemented in the Seurat framework \citep{hafemeister2019normalization}. After applying normalization, the distribution of normalized gene expression counts exhibited an approximately Gaussian-like shape.

\begin{figure}[ht!]
  \centering
  \includegraphics[scale=0.5]{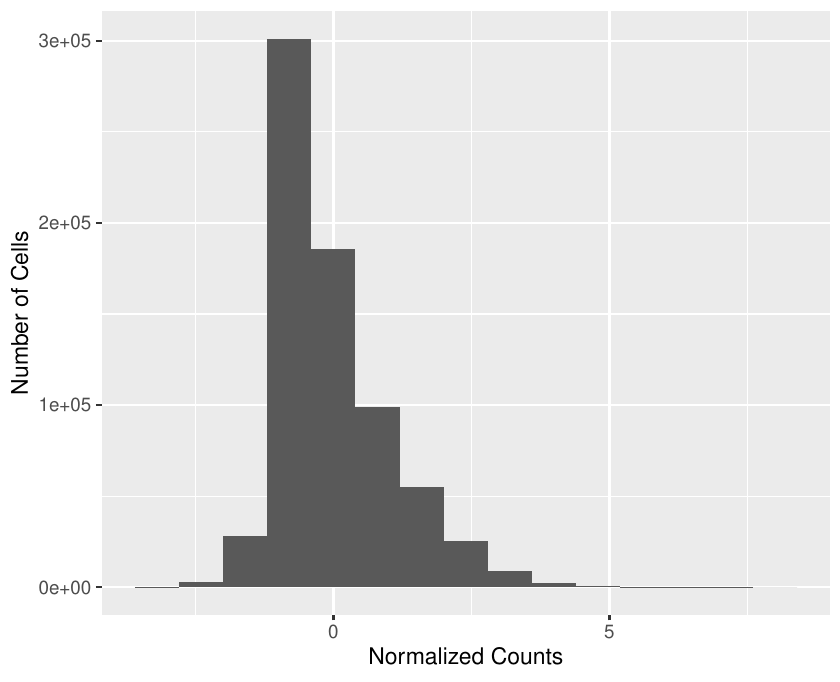}
  \caption{Distribution of normalized gene expression}
  \label{suppfig:histogram}
\end{figure}

\subsection{Incorporation of spatial coordinates at FOV-level:} We consider a power correlation matrix with $[k,k']$ element defined as $\rho^{d_{kk'}}$, where $d_{kk'}$ denotes the Euclidean distance between the centroids of FOVs $k$ and $k'$. To estimate $\rho$, we first aggregate cell-level expression data to the FOV level. Specifically, for each FOV we compute the mean expression of all EMT genes, which serves as the variable of interest for calculating spatial autocorrelation. Pairwise Euclidean distances between FOV centroids are then used to construct a Gaussian kernel spatial weight matrix with entries
\[
w_{kk'} = \exp\left(-\frac{d_{kk'}^2}{2\sigma^2}\right),
\]
where $\sigma = 1$ serves as the bandwidth parameter. The spatial autocorrelation parameter $\rho$ is quantified using Moran’s $I$ statistic, with the Gaussian kernel matrix serving as the spatial weight structure.

\paragraph{Incorporation of spatial coordinates at cell-level:} Prior to fitting the Gaussian process, the $x$ and $y$ coordinates of all cells are scaled to ensure comparable ranges and numerical stability.  

\subsection{Hyperparameters Specification} For the selection indicators, we specify the prior hyperparameters as  
$M = \mathbf{0}, U = I_{23}, V = \big[ V_{kk'} \big]_{23 \times 23}, 
V_{kk'} = \frac{1}{50} \rho^{\, d_{kk'}}$.
Here, $d_{kk'}$ denotes the Euclidean distance between the centroids of FOVs $k$ and $k'$, and $\rho^{d_{kk'}}$ defines the power correlation kernel applied elementwise.
The Gaussian process prior is controlled by shape parameters $a_{\mathrm{gp}} = 0.01$ and $b_{\mathrm{gp}} = 0.5$, with polynomial degree set to be 10. We set $a_{w} = 10$ and $b_{w} = 10$. Convergence is declared once the magnitude of the difference between updated and previous parameter estimates falls below $0.001$.

\subsection{Connectivity Between Genes in EMT Pathways}
We provide additional results characterizing both connectivity degree and gene expression patterns for genes that exhibit statistically significant differences in connectivity degree along the tumor progression gradient or are identified as hub genes. Specifically, we present gene-level connectivity summaries together with corresponding expression levels, highlighting genes whose network involvement varies markedly across spatial contexts. 

\begin{figure}[ht!]
  \centering
  \includegraphics[scale=0.5]{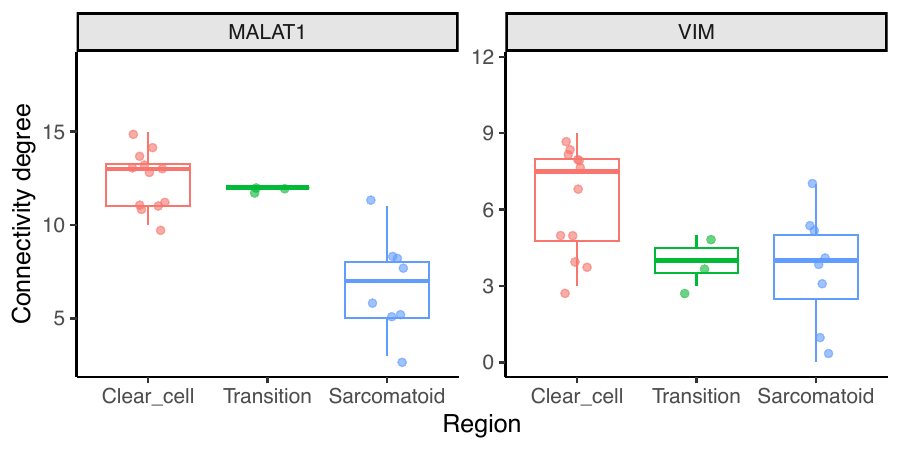}
  \caption{\textbf{MALAT1 and VIM as hub genes with high connectivity scores.} Boxplots show the distribution of FOV connectivity degree for MALAT1 and VIM across clear cell, transitional, and sarcomatoid regions. Connectivity degree is defined as the number of significant gene–gene associations incident to each gene within a given FOV. Each point represents one FOV.}
  \label{suppfig:hub_connectivity}
\end{figure}
\newpage
\begin{figure}[ht!]
  \centering
  \includegraphics[scale=0.5]{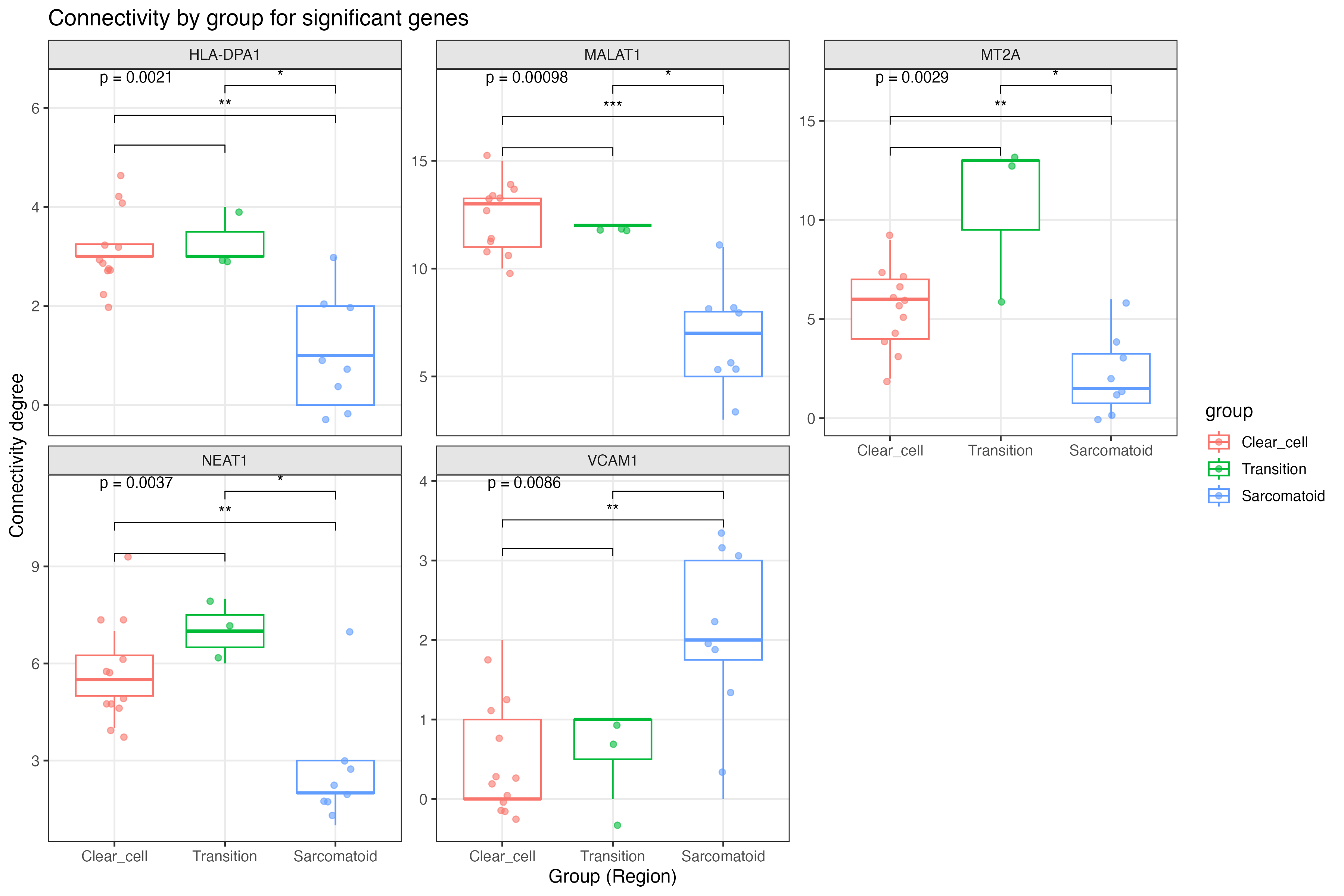}
  \caption{\textbf{Regional differences in gene connectivity for HLA-DPA1, MALAT1, MT2A, NEAT1 and VCAM1} Boxplots show the distribution of FOV connectivity degree for five genes (HLA-DPA1, MALAT1, MT2A, NEAT1 and VCAM1) which show significant regional differences after multiple testing correction. Connectivity degree is defined as the number of significant gene–gene associations incident to each gene within a given FOV. Each point represents one FOV. Overall differences across regions were assessed using the Kruskal–Wallis test, followed by pairwise Wilcoxon rank-sum tests with significance indicated by stars (* $p < 0.05$, ** $p < 0.01$, *** $p < 0.001$)}
  \label{suppfig:connecticity_difference}
\end{figure}
\newpage
\begin{figure}[H]
  \centering
  \includegraphics[scale=0.5]{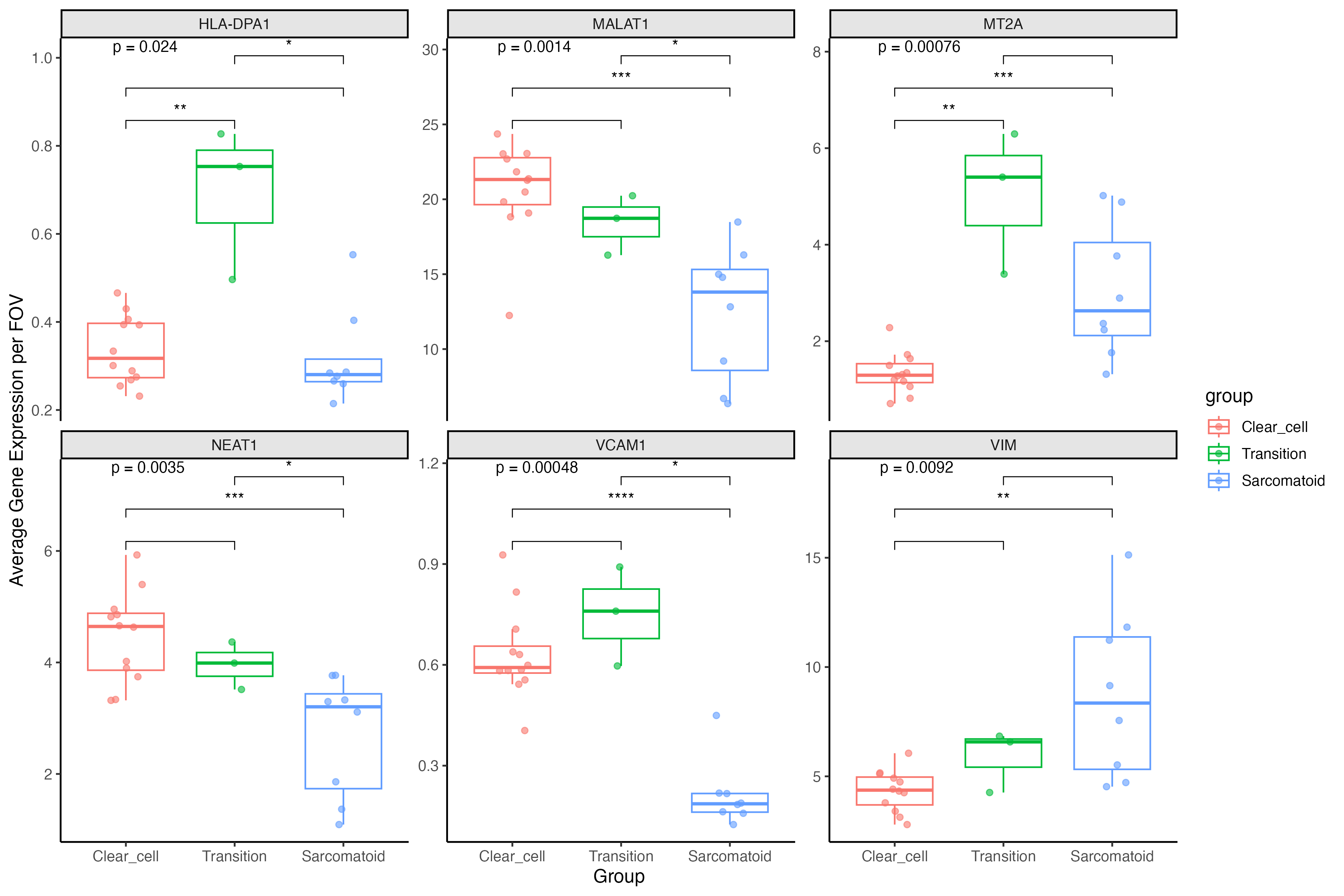}
  \caption{\textbf{Regional differences in gene expression for HLA-DPA1, MALAT1, MT2A, NEAT1, VCAM1 and VIM.} Boxplots display the distribution of mean gene expression at the FOV level for six genes (HLA-DPA1, MALAT1, MT2A, NEAT1, VCAM1 and VIM). Each point corresponds to a single FOV. Overall differences across regions were evaluated using the Kruskal--Wallis test, followed by pairwise Wilcoxon rank-sum tests, with statistical significance indicated by stars (* $p < 0.05$, ** $p < 0.01$, *** $p < 0.001$).}
  \label{suppfig:expression_difference}
\end{figure} 



\end{document}